\documentclass[iop,twocolappendix]{emulateapj}
\usepackage{natbib}
\usepackage{color}
\usepackage{mathptmx}
\usepackage{amsmath}
\usepackage{url}
\usepackage{multirow}
\usepackage{verbatim}
\usepackage[figuresright]{rotating}
\usepackage{afterpage}  
\usepackage{refcount}
\usepackage{threeparttable}
\usepackage{xcolor}
\def\lsim{\lower.5ex\hbox{$\; \buildrel < \over \sim \;$}}
\def\gsim{\lower.5ex\hbox{$\; \buildrel > \over \sim \;$}}

\def\apj{ApJ}
\def\apjs{ApJS}

\def\mnras{MNRAS}
\def\nat{Nature}

\def\apjl{ApJL}

\def\aap{Astronomy \& Astrophysics}
\def\aapr{Astronomy and Astrophysics Review}

\begin{document}

\shorttitle{Stellar Feedback and High-redshift Quasar Growth}
\shortauthors{J. Kim et al.}

\title{High-redshift Galaxy Formation with Self-consistently Modeled Stars and Massive Black Holes: Stellar Feedback and Quasar Growth}

\author{Ji-hoon Kim$^{1,2}$}
\author{John H. Wise$^{3}$}
\author{Tom Abel$^{4, 5}$}
\author{Yongseok Jo$^{1}$}
\author{Joel R. Primack$^{6}$}
\author{Philip F. Hopkins$^{7}$}

\affil{$^{1}$Center for Theoretical Physics, Department of Physics and Astronomy, Seoul National University, Seoul 08826, Korea}
\affil{$^{2}$\url{me@jihoonkim.org}}
\affil{$^{3}$Center for Relativistic Astrophysics, School of Physics, Georgia Institute of Technology, Atlanta, GA 30332, USA}
\affil{$^{4}$Kavli Institute for Particle Astrophysics and Cosmology, SLAC National Accelerator Laboratory, Menlo Park, CA 94025, USA}
\affil{$^{5}$Department of Physics, Stanford University, Stanford, CA 94305, USA}
\affil{$^{6}$Department of Physics, University of California at Santa Cruz, Santa Cruz, CA 95064, USA} 
\affil{$^{7}$TAPIR, Department of Astronomy, California Institute of Technology, Pasadena, CA 91125, USA}

\begin{abstract}
As computational resolution of modern cosmological simulations reach ever so close to resolving individual star-forming clumps in a galaxy, a need for ``resolution-appropriate'' physics for a galaxy-scale simulation has never been greater.  
To this end, we introduce a self-consistent numerical framework that includes explicit treatments of feedback from star-forming molecular clouds (SFMCs) and massive black holes (MBHs).  
In addition to the thermal supernovae feedback from SFMC particles, photoionizing radiation from both SFMCs and MBHs is tracked through full 3-dimensional ray tracing. 
A mechanical feedback channel from MBHs is also considered.  
Using our framework, we perform a state-of-the-art cosmological simulation of a quasar-host galaxy at $z\sim7.5$ for $\sim 25$ Myrs with all relevant galactic components such as dark matter, gas, SFMCs, and an embedded MBH seed of  $\gtrsim10^6\,\,{\rm M}_{\odot}$. 
We find that feedback from SFMCs and an accreting MBH suppresses runaway star formation locally in the galactic core region.
Newly included radiation feedback from SFMCs, combined with feedback from the MBH, helps the MBH grow faster by retaining gas that eventually accretes on to the MBH.  
Our experiment demonstrates that previously undiscussed types of interplay between gas, SFMCs, and a MBH may hold important clues about the growth and feedback of quasars and their host galaxies in the high-redshift Universe.
\end{abstract}

\keywords{galaxies: formation -- galaxies: evolution -- galaxies: kinematics and dynamics -- galaxies: star formation -- galaxies: nuclei -- ISM: structure -- stars: formation -- quasars: supermassive black holes -- methods: numerical}

\section{INTRODUCTION}\label{intro}

The very massive black holes lurking at the centers of many large galaxies in the Universe have been the topic of hard-working observers and theorists over the past decades.
It has now become a consensus view that these massive black holes (MBHs) and their host galaxies have grown together under each other's influence.
Moreover, observations indicate that extremely massive black holes of mass $\gtrsim10^9\,\, {\rm M}_{\odot}$ started to exist in the $z>7$ era, only a few hundred Myrs after the Big Bang \citep[e.g.,][]{2011Natur.474..616M, 2018Natur.553..473B}, and many more prevailed in the $z\sim6$ era \citep[e.g.,][]{2013ApJ...779...24V, 2015Natur.518..512W}.
The discovery of surprisingly massive black holes in the early Universe inevitably raises intriguing questions about our understanding of the Universe:  
How did MBHs acquire such large masses in such a short time? 
Do we need any new physical mechanism to explain the rapid growth of MBHs?
How do MBHs interact with the interstellar medium (ISM) of their host galaxies? 
Do high-$z$ MBHs imply that the age of the Universe estimated by the standard $\Lambda$CDM cosmology is inaccurate after all?
The evolution of MBHs in the early Universe is thus considered as an ultimate testing ground for our contemporary understandings of astrophysics, black hole physics, and cosmology.  

A commonly referred scenario for a high-$z$ MBH formation begins with a remnant black hole of tens of solar masses that forms when a first generation Population III star dies at $z \gtrsim 10$. 
Then a sequence of galaxy mergers and subsequently, merging of their embedded black holes are thought to have led to a $\gtrsim10^9\,\, {\rm M}_{\odot}$ MBH by $z > 7$ \citep[e.g.,][see also other scenarios with more massive seed black holes such as direct collapse black holes --- to be discussed in Section \ref{stability}]{2001ApJ...552..459H, 2010A&ARv..18..279V}.
A numerical evaluation of this hypothesis requires us to implement {\it self-consistent} physics models surrounding the evolution of MBHs in simulations:  
How is the interstellar gas consumed through star formation and MBH accretion? 
How do radiation and winds from the MBHs in turn self-regulate the growth of their own and their host galaxies? 
How do young stars' radiation and supernovae explosions curb subsequent star formation and keep the interstellar gas from being hastily consumed?  
An unabridged, self-consistent modeling of how galactic ingredients --- gas, stars, and MBHs --- interact with one another at different scales is indeed crucial to numerically probe the evolution of MBHs.  

Aided by the advances in parallel computing, modern cosmological simulations with large computational domains (of at least $\gtrsim 10$ Mpc) are now starting to resolve structures as small as individual star-forming clouds (of $\lesssim 10$ pc) inside a galaxy \citep[e.g.,][]{2018MNRAS.480.4842C, 2018MNRAS.480..800H, 2018MNRAS.474.4232K}.  
This type of high-resolution simulation is an ideal vehicle to investigate how a high-$z$ MBH grew as they resolve gas inflows from large to small scales simultaneously, and allows us to sample one or more sufficiently massive MBH seed. 
Such simulations will be indispensable to fully grasp how small-scale physics of MBHs is so tightly linked with overall galactic evolution and morphology. 
However, note that even a simulation with the best numerical resolution would not be useful to its full potential unless it is accompanied by self-consistent physics models that are {\it appropriate at that particular resolution}.  
In fact, a high-resolution simulation without ``resolution-appropriate'' physics models could very well produce meaningless, if not inaccurate, results.  
As an example, for a kpc-resolution simulation, a simple stellar feedback prescription that describes only supernovae explosions is likely sufficient. 
But for a pc-resolution simulation with a corresponding timestep of $\Delta\, t \sim 3 \times 10^4$ yr at $10^4$ K, a better feedback prescription is required that includes early channels such as photoionizing radiation from young stars  \citep[ages of $\lesssim 3 \,\,{\rm Myr} \simeq 100\, \Delta\, t$; see e.g.,][]{2013MNRAS.428..129S, 2013ApJ...775..109K, 2018MNRAS.480..800H}.
Without a temporally-resolved stellar feedback model appropriate for the resolution, even the pc-resolution simulation would not be useful to its full potential.  

Despite its best recent efforts \citep[e.g.,][]{2010MNRAS.407.1529H, 2015MNRAS.452..575S, 2017MNRAS.466.3331D, 2017MNRAS.465.3291W, 2018MNRAS.480..800H, 2018MNRAS.tmp.1353T}, the numerical galaxy-MBH formation community is yet to converge on ``resolution-appropriate'' physics models that are compatible with the best resolution in the field ($\lesssim 10$ pc).  
Instead, previous galaxy-MBH formation simulations have sometimes had recourse to empirically-tuned sub-resolution recipes, such as boosted Bondi accretion estimates, or bimodal feedback channels --- quasar- and radio-mode --- for MBHs.  
While these prescriptions have been instrumental in calculations with $\sim100$  pc resolution, the community is rightfully searching for better treatments for sub-resolution physics  that do not add too many tunable parameters to an already complex problem.    
Only when a self-consistent framework is implemented that appropriately describes the physical processes between gas, stars, and MBHs at a given resolution can we acquire any insight from the simulation about high-$z$ MBHs.  

In this work, we introduce a new breed of simulation that is aimed at resolving the ``mismatch'' between the best resolution in the field and the physics models used. 

(1) First, for MBH physics, we improve the framework first introduced in  \cite{2011ApJ...738...54K}.  
Our framework describes how the interstellar gas flows around and accretes onto a MBH using high-resolution adaptive mesh refinement (AMR).  
Rather than resorting only to a thermal feedback prescription, we employ two channels of MBH feedback: radiative feedback --- photoionizing radiation traced via full 3-dimensional adaptive ray tracing --- and mechanical feedback --- bipolar winds resolved in AMR (Section \ref{method-mbh}).   
Earlier or similar versions of this framework have been used to study co-evolution of a star-forming galaxy and its embedded MBH \citep{2011ApJ...738...54K}, evolution of MBHs in merging galaxies \citep{Kim2011Dissertation}, and the early growth of a MBH seed at $ 8 \lesssim z \lesssim 15$ \citep{2014ApJ...797..139A, 2018MNRAS.476.5016L, 2018ApJ...865..126S}, sometimes at very high resolution depicting the near-relativistic jets from a MBH seed accreting at super-Eddington rates \citep{2019MNRAS.486.3892R}.
In particular, in \cite{2011ApJ...738...54K} our MBH framework proved rewarding in depicting a new mode of MBH feedback that locally suppresses star formation in the galaxy's core, and self-regulates its own growth by heating up the surrounding ISM.    

(2) Second, for the physics of star-forming molecular clouds (SFMCs), we adopt and improve the framework established in \cite{2013ApJ...775..109K, 2013ApJ...779....8K}. 
Here, the framework describes radiation from each of $\sim10^4$ SFMC particles --- each of which represents a mass of $\gtrsim 10^3\,\,{\rm M}_{\odot}$ --- by tracing the UV photons on the fly, as well as thermal supernovae feedback  (Section \ref{method-sfmc}).  
This means that our approach considers the early stellar feedback channels before the supernovae explode.  
Joined with high spatial resolution, this framework was successfully applied to a dwarf-sized galaxy to study how stellar radiation escapes SFMCs and the galaxy \citep{2013ApJ...775..109K}, and how different life stages of SFMCs manifest themselves in a spatially-resolved star formation relation between H$\alpha$ and H$_2$ surface densities \citep{2013ApJ...779....8K}. 

Using our framework aimed at self-consistently modeling MBH and SFMC physics, in this paper we study the interactions between different galactic ingredients of a high-$z$ galaxy  with high spatial resolution. 
With a MBH of mass $\gtrsim 10^6\,\,{\rm M}_{\odot}$ embedded in a $\sim7 \times10^{10}\,\,{\rm M}_{\odot}$ halo at $z\sim7.5$, we study a massive quasar-host --- assuming its MBH has already experienced a sizable early buildup (from $\sim 10^2\,\,{\rm M}_{\odot}$ to $\sim 10^6\,\,{\rm M}_{\odot}$).
The novelty of our physics models means that they can shed light on aspects of galaxy-MBH evolution that previous approaches could not, and provide unique perspectives for the high-$z$ MBH puzzle. 
An example would be how interstellar gas is consumed by two competing channels, star formation and MBH accretion, in the vicinity of a MBH.  
As the discussion in the present paper will make clear, we found that radiation from young stars, combined with radiation and winds from the MBH, helps retain interstellar gas near the MBH which might otherwise have been rapidly consumed by runaway star formation.  
The unused gas is then available as a fuel for the MBH, increasing its growth rate when compared with the run without stellar radiation feedback (see Section \ref{results-mbh} for detailed discussion).  
This example demonstrates that the interplay between gas, stars, and MBHs that were never realized in previous simulations may contain critical information about the expeditious growth of MBHs at high $z$.    

The remainder of the paper is organized as follows.  
In Section \ref{methodology} we briefly explain the simulation framework, the physics models, and the initial condition of the experiment.  
Next, Sections \ref{results} and \ref{discussion} are devoted to the results and discussion of the reported simulations, with particular emphases on the interplay between gas, SFMCs, and a MBH.   
Finally in Section \ref{conclusion} we summarize our findings and implications, with discussions on the future direction of the project.

%\hspace{5mm}

\section{METHODOLOGY}\label{methodology}

Our numerical experiment is designed to depict a galactic halo of $\sim 7\times 10^{10}\,\,{\rm M}_{\odot}$ at $z\sim7.5$ with an embedded MBH seed of $\gtrsim10^6\,\,{\rm M}_{\odot}$. 
The simulations presented here were run with the {\sc Enzo} code \citep{1997ASPC..123..363B, 1999ASSL..240...19N, 2007arXiv0705.1556N, 2014ApJS..211...19B_short}.\footnote{The website is http://www.enzo-project.org/.} 
Our variant of {\sc Enzo} includes all physics relevant in a galaxy-scale simulation as well as improved versions of MBH physics and SFMC physics modules (improved from \citealt{2011ApJ...738...54K} and \citealt{2013ApJ...775..109K, 2013ApJ...779....8K}, respectively).  
For completeness and the reproducibility of the experiment, we review the framework's key features  focusing on the improvements  when compared with our earlier studies.  

\subsection{Hydrodynamics, Refinement, and Cooling}\label{method-hydro}

We employ the ZEUS hydrodynamics solver to evolve the collisional fluid \citep{1992ApJS...80..753S, 1992ApJS...80..791S}.\footnote{We find that the typical runtime parameter combinations of the PPM solver in {\sc Enzo}  often stall or fall back to the lower-order-accurate solvers, especially when we combine the on-the-fly calculation of radiative feedback from SFMC and MBH particles, and the wind feedback from MBH particles.  Thus, in order to keep the consistency in the numerical accuracy we achieved, we employ a less-accurate, but more stable ZEUS solver.  We however note that the more sophisticated subgrid physics we can include with ZEUS could justify our choice.}
The mass thresholds for gas and particles above which a cell is refined depend on a refinement level {\it l} as
\begin{align}
M_{\rm ref, \,gas}^{l} &= 2^{-0.446\, l} \times4\times(1/2^3)^5\times \Omega_{\rm b} \rho_0 \Delta x_0^3\\
M_{\rm ref, \,part}^{l} &= 2^{-0.143\, l} \times4\times(1/2^3)^5\times \Omega_{\rm m} \rho_0 \Delta x_0^3
\end{align}
where $\Delta x_0$ is the cell size at a root level, and $\rho_0 = 3 H^2 / 8 \pi G$ is the critical density of the Universe. 
A factor $(1/2^3)^5$ is to refine all the cells in the first five nested levels.  
The $\Lambda$CDM cosmology we adopt matches the one used to set up the initial condition (see Section \ref{method-IC}) and is consistent with {\it WMAP7+SNe+BAO}: 
$\Omega_{\rm m} = 0.272$, $\Omega_{\rm b} = 0.0455$, and $H = 70.2 \,\,{\rm km\,\, s}^{-1} {\rm Mpc}^{-1}$ \citep{2011ApJS..192...18K_short}.  
Readers may notice that our choice of $M_{\rm ref}^{l}$ refines the grids more on small scales (``super-Lagrangian'').
With a root resolution of $128^3$ in a $(60 \,\,\,{\rm comoving} \,\,h^{-1}\,{\rm Mpc})^3$ box, the finest mesh at $l=14$ gives a maximal physical resolution $\Delta x_{14} = 4.79\,\,{\rm pc}$ at $z=7.5$, approximately in accord with the Jeans length for a dense gas clump of $n=10^3$ ${\rm cm^{-3}}$ at $\sim$100 K.   
The corresponding Jeans mass, $2000 \,\,{\rm M}_{\odot}$, is then used as a threshold for star formation at $l=14$, above which the gas cell collapses to spawn a SFMC particle  (see Section \ref{method-sfmc} and \citealt{2013ApJ...775..109K} for more discussion).
$M_{\rm ref, \,gas}^{14} = 3000 \,\,{\rm M}_{\odot} $ is chosen so that it is consistent with the above Jeans argument, while $M_{\rm ref, \,part}^{14} = 1.13\times10^6 \,\,{\rm M}_{\odot} $ is set to the mass of four finest dark matter particles.

Non-equilibrium chemistry module in {\sc Enzo} tracks six species (H, ${\rm H}^+$, He, ${\rm He}^+$, ${\rm He}^{++}$, ${\rm e}^-$) and six collisional processes between them.  
Meanwhile, {\sc Enzo} computes the primordial cooling rate by considering collisional excitation/ionization cooling, recombination cooling, Bremsstrahlung cooling, and CMB Compton cooling for hydrogen and helium \citep{1997NewA....2..209A}. 
When above $10^4$ K, added to the primordial rate is the metal cooling rate, $\Delta \Lambda(Z) = \Lambda_{\rm net}(Z) - \Lambda_{\rm net}(0)$,  where $\Lambda_{\rm net}$ is the net cooling rate tabulated in \cite{1993ApJS...88..253S}.
If below $10^4$ K, we use an approximated $\Lambda(T)$ found in \cite{2002ApJ...564L..97K} with corrections noted in \cite{2010ApJ...715.1302V}.
Unlike \cite{2011ApJ...738...54K}, metagalactic UV background radiation is not included as we consider one of the first, massive galaxies at $z\gtrsim7.5$ that generates its own cosmological HII region and is unaffected by radiation from other galaxies.

\begin{figure*}
\begin{center}
\includegraphics[width=1\textwidth]{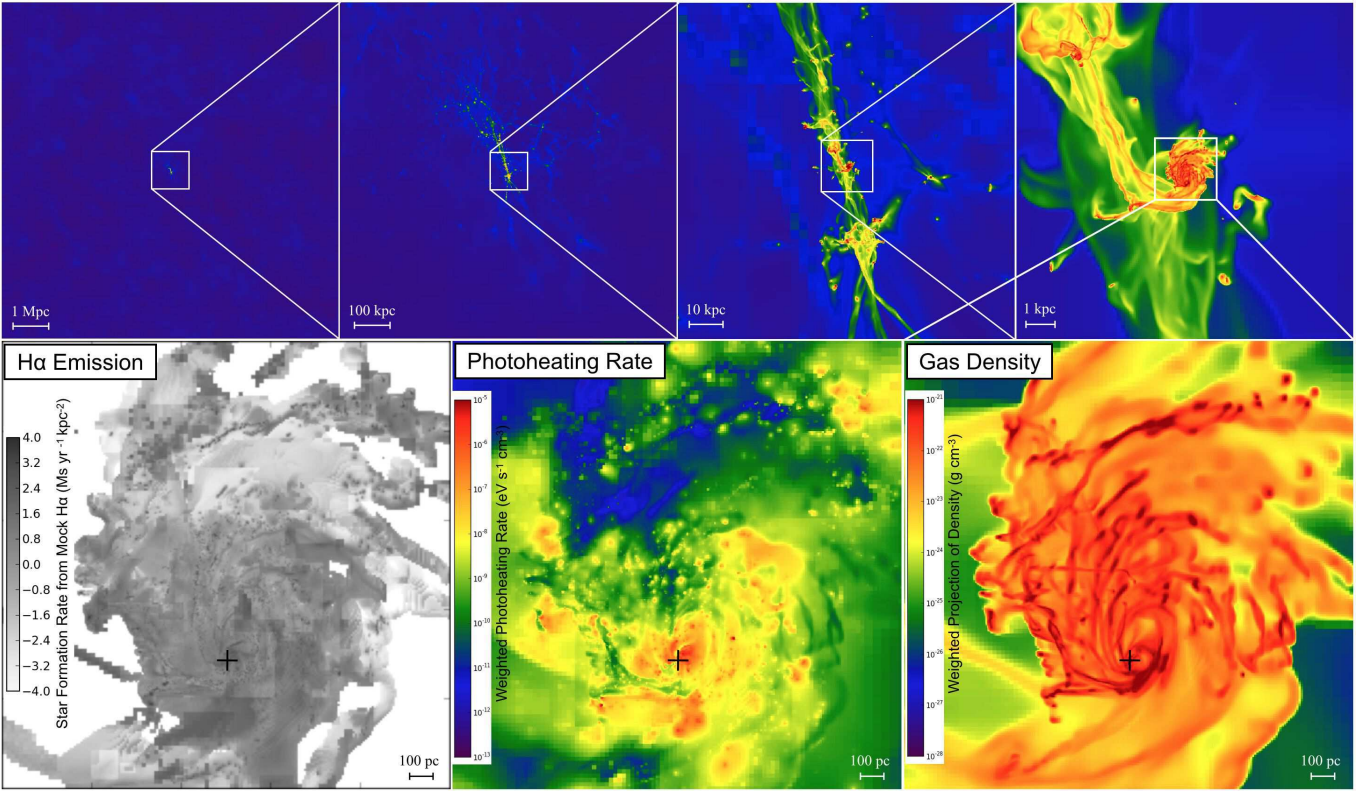}
    \caption{Overview of the target galaxy in a halo of total mass $\sim 7\times 10^{10}\,\,{\rm M}_{\odot}$ in our experiment at $z=7.65$.   {\it Clockwise from top left to bottom right:} projected gas densities at large to small scales seen from a face-on angle.   {\it Bottom middle:} projected photo-heating rate density calculated by ray-tracing the UV photons from radiating SFMCs.   {\it Bottom left:} star formation rate surface density estimated from mock H$\alpha$ emission \citep{2013ApJ...779....8K}.  This halo eventually grows into a $\sim 10^{13}\,\,{\rm M}_{\odot}$ group at $z=0$ (see Section \ref{method-IC}). The high-resolution, full color version of this figure is available online and at http://www.jihoonkim.org/.
\label{fig:overview}}
\end{center}
\vspace{2pt}
\end{figure*}

\subsection{Star-forming Molecular Cloud (SFMC) Physics}\label{method-sfmc}

Now we summarize our SFMC physics model, improved from a version tested in \cite{2013ApJ...775..109K, 2013ApJ...779....8K}.    
The finest gas cell of size $\Delta \,x_{14}$ and density $\rho_{\rm gas}$  produces a SFMC particle of initial mass $M_{\rm MC}^{\rm init} = \epsilon_{\star} \rho_{\rm gas} \Delta \,x_{14}^3$ with efficiency $\epsilon_{\star} = 0.5$ \citep[a value established and extensively tested in connection with the other star formation criteria in earlier studies; see][]{2011ApJ...738...54K, 2013ApJ...775..109K, 2013ApJ...779....8K} when 
{\it (a)} the proton number density exceeds the threshold $n_{\rm thres} = 10^3$ ${\rm cm^{-3}}$, 
{\it (b)} the velocity flow is converging, 
{\it (c)} the cell's cooling time is shorter than the gas dynamical time, and 
{\it (d)} the particle produced has at least $M_{\rm thres} = 1000 \,\,{\rm M}_{\odot}$.
Because the gas is instantly converted to a particle when a gas cell of $M_{\rm thres} / \epsilon_{\star}= 2000 \,\,{\rm M}_{\odot}$ becomes Jeans unstable, the gas mass in the finest cell never reaches the refinement threshold $M_{\rm ref, \,gas}^{14} = 3000 \,\,{\rm M}_{\odot}$ described in Section \ref{method-hydro}, ensuring the consistency between the SFMC formation machinery and the refinement criteria (Section \ref{method-hydro} and \citealt{2013ApJ...775..109K} for more discussion).
The SFMC particle thus represents a self-gravitating star-forming cloud decoupled from the rest of gas.\footnote{Readers should note that SFMC particles could form even in the cell where the MBH sits.  See Section \ref{results-mbh} for related discussion.}

Once created, a SFMC particle describes feedback from a population of stars in it, with its mass evolving as
\begin{equation}
M_{\rm MC}(t) = 
\begin{cases}
M_{\rm MC}^{\rm init}  &\,\,\,\, {\rm if}\,\,\,\, T/{\rm Myr} <4 \\[5pt]
M_{\rm MC}^{\rm init}   \left[1 - 0.8 \left( {T- 4} \over 40-4 \right) \right] &\,\,\,\, {\rm if}\,\,\,\, 4<T/{\rm Myr} <40 \\[5pt]
0.2\,M_{\rm MC}^{\rm init}  &\,\,\,\, {\rm if}\,\,\,\, T/{\rm Myr}  >40 
\end{cases}
\label{eq:SFMC_mass}
\end{equation}
where $T = t-t_{\rm cr}$ is the particle age in Myr with particle creation time $t_{\rm cr}$.  
The above formulation entails different life stages of a SFMC particle: 
{\it (a)} At its birth, $0.24\,\,M_{\rm MC}^{\rm init}$ is considered to have instantly turned into stars.
The rest, $0.76\,\,M_{\rm MC}^{\rm init}$, never participates in star formation, modeling inefficient star formation in molecular clouds \citep[e.g.,][]{2007ApJ...654..304K, 2010ApJ...709..191M}.  
{\it (b)} Among the initial stellar mass of $0.24\,\,M_{\rm MC}^{\rm init}$, 83\% ends up forever locked in the particle as permanent stellar mass, $M_{\star} = 0.20\,\,M_{\rm MC}^{\rm init}$.
{\it (c)} The remaining 17\%, or $0.04\,\,M_{\rm MC}^{\rm init}$, represents stars massive enough to ignite  Type II SNe that continuously inject thermal energy from $T=4$ to 40 Myr.\footnote{$0.04\,\,M_{\rm MC}^{\rm init}$ corresponds to 17\% of the initial stellar mass assuming that massive stars with $> 8\,\,{\rm M}_{\odot}$ commence Type II SNe in a \cite{1955ApJ...121..161S} initial mass function between 0.1 and $300\,\,{\rm M}_{\odot}$.  Note that the mass evolution Eq.(\ref{eq:SFMC_mass}) and the duration of SNe are updated from \cite{2013ApJ...775..109K}.}
During this time, $M_{\rm MC}^{\rm init} - M_{\star} = 0.8\,\,M_{\rm MC}^{\rm init}$ is gradually released back to the ISM, along with $3.75 \times 10^{-5}$ of the rest mass energy of $M_{\star}$ in thermal form.
This energy corresponds to $2\times10^{51}$ ergs per every 30 ${\rm M}_{\odot}$ of permanent stellar mass $M_{\star}$ produced.  
2\% of the ejecta mass is considered as metals.
It is noted that the energy and duration of our thermal feedback model updates those in \cite{2011ApJ...738...54K} or \cite{2013ApJ...775..109K, 2013ApJ...779....8K}, and roughly matches previous studies such as \citet[based on {\sc Starburst 99} estimation]{2009ApJ...695..292C}.

In addition to thermal supernovae, SFMC particles with ages $T < 100$ Myr may heat up the surrounding ISM by emitting UV photons.  
Using {\sc Enzo}'s radiative transfer machinery described in previous work \citep[e.g.,][]{2002MNRAS.330L..53A, 2011MNRAS.414.3458W} as well as in \cite{2011ApJ...738...54K} and \cite{2013ApJ...775..109K, 2013ApJ...779....8K}, we perform an explicit 3-dimensional ray tracing calculation to evolve the radiation fields throughout the galaxy. 
The radiation luminosity is assigned to each SFMC particle by
\begin{align}
L_{\rm MC}(t) = q_{\rm MC}  \, E_{\rm ph}  \, M_{\rm MC}(t)  
\label{eq:lum_MC}
\end{align}
where $q_{\rm MC} = 6.3 \times 10^{46} \,\,{\rm photons} \,\, {\rm s}^{-1} M^{-1}_{\odot}$ is the lifetime-averaged ionizing luminosity \citep{2010ApJ...709..424M},  and $E_{\rm ph} = 16.0\,\, {\rm eV}$ is the mean monochromatic energy per photon \citep{2006ApJS..162..281W}.\footnote{Our choice to use $M_{\rm MC}(i,t)$ in estimating $L_{\rm MC}(t)$ rather than the stellar mass is to approximately compensate for various other early channels of SFMC feedback beyond photoionization, such as protostellar outflows and stellar winds.  See \cite{2013ApJ...775..109K} for complete discussion.}
Initially, $12\times4^3$ rays are isotropically cast from the particle with $ L_{\rm MC}(t) \, dt_{\rm ph} / (E_{\rm ph} \cdot 12 \times 4^3) $ photons per ray. 
Here the radiation timestep $dt_{\rm ph}$ is adaptively set by the code, comparable with the finest hydrodynamic timestep most of the time \citep{2011MNRAS.414.3458W}. 
Each ray is split into four child rays whenever the area associated with a ray becomes larger than $0.2\, (\Delta x)^2$ of a local cell, and is traced until the edge of the computational domain.\footnote{To speed up the calculation multiple measures are placed in the ray tracing machinery: 
{\it (a)} The photon luminosities $L_{\rm MC}(t)$ are assigned only to the SFMCs that are within the virial radius of the target galaxy (15.1 proper kpc from the galactic center at $z\sim 7.5$, or 128.2 comoving kpc; see Section \ref{method-IC}). 
{\it (b)} Two SFMC particles are merged if separated by less than 16 proper pc in space and 0.1 Myr in creation time, in a way that conserves mass and momentum.  
{\it (c)} A ray is no longer split when it is more than 3 kpc from the source.
{\it (d)} Two rays are merged if distances from the sources are more than 8 times the separation between the ray sources.  \label{footnote:rt}}     
Photons in the ray affect the surrounding ISM, first by ionizing hydrogen with the rate of
\begin{align}
k_{\rm ph, H} &= { P_{\,\,\rm in} (1-e^{-\tau_{\rm H}}) \over n_{\rm H} (\Delta x)^3 dt_{\rm ph} }
\label{eq:k_ph}
\end{align}
where $P_{\,\,\rm in}$ is the number of incoming photons and $\tau_{\rm H} = n_{\rm H} \sigma_{\rm H}(E_{\rm ph}) dl$ is the optical depth of a cell; and second by heating the gas with the excess energy above the ionization threshold with the rate of
\begin{align}
\Gamma_{\rm H} &= k_{\rm ph, H} (E_{\rm ph} - 13.6\,\, {\rm eV}).
\label{eq:Gamma_H}
\end{align}

UV radiation is the only feedback channel before the thermal supernovae feedback kicks in at $T=4$ Myr.  
It is worth noting two points about it:  
{\it (a)} Although both hydrogen and helium are tracked in the chemistry module, only hydrogen is considered in photoionization and photoheating calculations \citep[as in][]{2013ApJ...775..109K}.
{\it (b)} We do not consider radiation pressure on gas --- either by hydrogen-ionizing UV photons \citep[unlike][]{2013ApJ...775..109K}, or by multiply-scattered IR photons \citep[e.g.,][]{2014MNRAS.445..581H, 2018MNRAS.480..800H}.  
Including the radiation pressure might strengthen the SFMC feedback, though controversies exist about the exact level of enhancement  \citep[e.g.,][]{2013MNRAS.434.2329K}.

\begin{figure*}
\begin{center}
\includegraphics[width=0.66\textwidth]{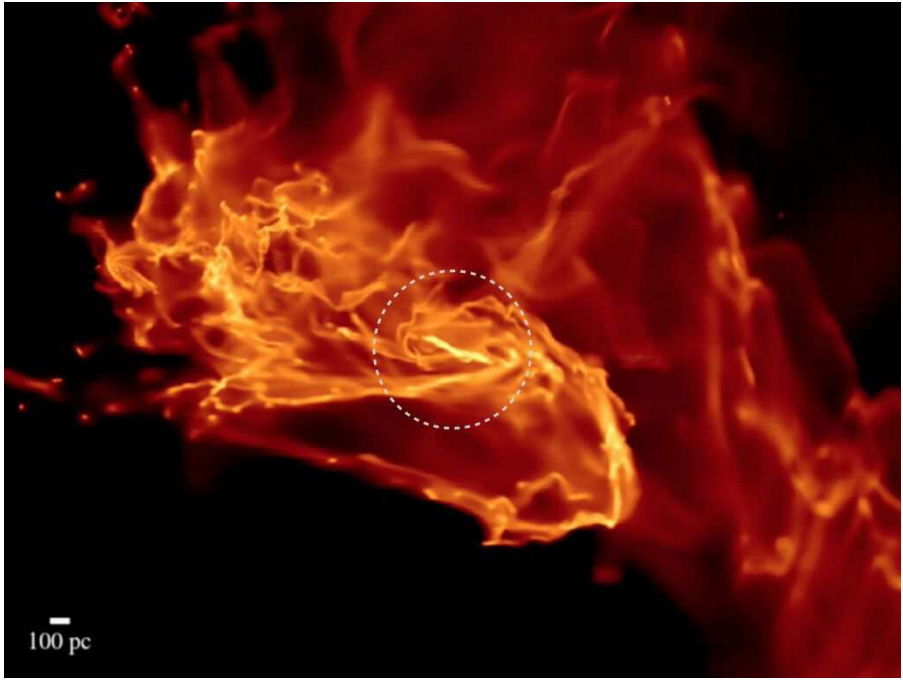}
\vspace{8pt}
    \caption{A snapshot of the target galaxy at $z=7.3$ ({\it Sim-SRTF+BH}). The white dashed circle in the middle denotes the target galaxy’s inner disk (seen from an edge-on angle; see also Figure \ref{fig:overview_perp}) harboring a central MBH.  For more information about the target galaxy, see Section \ref{method-IC} or the caption of Figure \ref{fig:overview}.
\label{fig:overview_large}}
\end{center}
\vspace{4pt}
\end{figure*}

\subsection{Massive Black Hole (MBH) Physics}\label{method-mbh}

We plant a MBH seed at the target galaxy's center as a source of gas accretion and feedback.  
The rate of accretion is estimated with the Bondi-Hoyle formula \citep{1952MNRAS.112..195B} as 
\begin{align}
\dot M_{\rm BH} =  {4 \pi G^2 M^2_{\rm BH} \rho_{\rm B}  \over  { c^3_{\rm s} }  }  
\label{eq:Mdot}
\end{align}
where $M_{\rm BH}$ is the mass of the MBH, $c_{\rm s}$ is the gas sound speed of the cell in which the MBH resides, and $\rho_{\rm B}$ is the density at the Bondi radius $R_{\rm B}$.
When $R_{\rm B}  = 2 G M_{\rm BH} / c^2_{\rm s} \simeq 86 \,\,{\rm pc} \,\,( M_{\rm BH} / 10^6 \,{\rm M}_{\odot} ) (  c_{\rm s} / 10 \,\,{\rm km\,s^{-1}}   )^{-2}$ is resolved with $\Delta \,x_{14}\, ( = 4.79 \,\,{\rm pc} \,\,\,{\rm at}\,\,\, z=7.5)$ around the MBH,\footnote{Our MBH is almost always sitting at the finest refinement level with a cell width $\Delta \,x_{14}$, as it is located at the peak of gas distribution near the galactic center.  This allows us to resolve the gas flow around the MBH with our best resolution, without necessarily requiring the cells around the MBH to be successively refined down to the finest level \citep[as in][]{2011ApJ...738...54K}.} $\rho_{\rm B}$ is extrapolated from the density $\rho_{\rm gas}$ of the cell the MBH is located in, as
\begin{align}
\rho_{\rm B} = \rho_{\rm gas} \,\cdot\, {\tt min} \left[ \left( {R_{\rm B} \over \Delta x_{14}} \right)^{-1.5}, \,\,1.0 \,\right]
\end{align}
following  \cite{2011ApJ...738...54K} and \cite{2010ApJ...709...27W}.  
The gas accreting onto the MBH is subtracted uniformly from the cells within $R_{\rm B}$.  
Notice that in Eq.(\ref{eq:Mdot}) we do not include an empirical boost factor (similar to \citealt{2011ApJ...738...54K} and other recent efforts by e.g., \citealt{2017MNRAS.465.3291W} and \citealt{2018MNRAS.tmp.1353T}).
Nor do we impose the Eddington accretion limit (similar to other previous studies by e.g., \citealt{2016MNRAS.456.2993L}).  
These two considerations  are often critical --- and controversial --- in estimating the growth rate of a MBH in a cosmological time scale.  
But not so much so in the present study in which our focus is to examine the interaction of a MBH and its host for a relatively short time (for $\sim$ 25 Myr; see Section \ref{method-IC}), not the {\it absolute} value of the MBH mass increase.\footnote{Surely, the Bondi accretion estimate will needs to be reconsidered in an even higher-resolution simulation.  See Section \ref{future} for more discussion.}
By the same token, our choice of a MBH seed mass $1.8\times10^6\,\,{\rm M}_{\odot}$ is not very crucial either, but simply designed to induce substantial gas accretion --- and feedback thereafter --- by the MBH.  
This is in line with the argument for a $8\times10^5\,\,h^{-1}{\rm M}_{\odot}$ black hole seed adopted in \cite{2017MNRAS.465.3291W} (see more discussion in Section \ref{method-IC}).

Once starting to accrete gas from its neighborhood, the total feedback energy rate by the MBH particle is given as  
\begin{align}
L_{\rm BH}(t) = \epsilon_{\rm r} \dot M_{\rm BH}(t) c^2 
\label{eq:lum_BH}
\end{align}
where $\epsilon_{\rm r}=0.1$ is the conversion factor from the rest mass energy of accreting gas to the MBH's feedback energy \citep{1973A&A....24..337S}. 
In the reported experiment, two channels of MBH feedback are implemented: radiative and mechanical. 
And each channel contributes equally (i.e., $0.5\,L_{\rm BH}(t)$ each) to the total feedback energy.  
To begin with, for the transfer calculation of radiation from the MBH, we assume that the UV luminosity of the MBH is proportional to its bolometric luminosity as 
\begin{align}
L_{\rm BH,\, UV}(t) = 0.5\,f_{\rm UV} L_{\rm BH}(t) 
\label{eq:lum_UV_BH}
\end{align}
with a proportionality coefficient $f_{\rm UV} = 0.1$. 
Our choice of $f_{\rm UV}$ is broadly consistent with the estimate given in \cite{2017MNRAS.464.1854B} using the \cite{2004MNRAS.347..144S} active galactic nuclei (AGN) spectrum \citep[e.g., $f_{\rm UV1} = 0.079$ in][]{2018MNRAS.tmp.1353T}.  
As in the case of SFMC photons (Section \ref{method-sfmc}), monochromatic energy of $E_{\rm ph} = 16.0\,\, {\rm eV}$ is used for photon rays.
Note that our $E_{\rm ph}$  is in line with the luminosity-weighted average in the energy band of the \cite{2004MNRAS.347..144S} spectrum that contributes to the ionization of neutral hydrogen \citep[$E_{\rm ph} = 18.0\,\, {\rm eV}$ in][]{2018MNRAS.tmp.1353T}.\footnote{Therefore, our formulation implies that the ISM surrounding the MBH is either optically thin to the photons in other energy bands (e.g., when $E \gg 16.0\,\, {\rm eV}$) or negligibly affected by the them (e.g., when $E \ll 16.0\,\, {\rm eV}$).   We implicitly assume that only the UV portion of the AGN spectrum can thermally couple with the surrounding ISM.} 
When the rays travel through the gas cells, the photons interacts with hydrogens via photoionization and photoheating, Eqs.(\ref{eq:k_ph}) and (\ref{eq:Gamma_H}), respecitvely.  
As in SFMC photons, we neglect radiation pressure by MBH photons, which is conservative since its inclusion would have furthered the feedback effect  \citep[e.g.,][]{2011MNRAS.412.1341D, 2017MNRAS.464.1854B, 2018MNRAS.473.4197C}.   

\begin{table*}  %remove * for the table fitting in columns
\caption[Simulation suite description]{Simulation Suite Description}
\centering     
\begin{threeparttable}
\begin{tabular}{l  l  ||  c  c  c  c } 
\hline\hline   
\multicolumn{2}{c ||}{Physics\tnote{\textdagger}} & $\,\,\,\,\,\,\,\,\,${\it Sim-SRTF+BH}   & $\,\,\,\,\,${\it Sim-STF+BH}  & $\,\,\,\,\,${\it Sim-SRTF}  \\ [1ex] 
\hline      
SFMC formation\tnote{\textdaggerdbl}     & (Section \ref{method-sfmc}) &$\,\,\,\,\,\,\,\,\,$\textcircled{}&$\,\,\,\,\,$\textcircled{}&$\,\,\,\,\,$\textcircled{}\\    
SFMC radiative feedback -- ionizing radiation  & (Section \ref{method-sfmc}) &$\,\,\,\,\,\,\,\,\,$\textcircled{}&$\,\,\,\,\,$$\times$&$\,\,\,\,\,$\textcircled{}\\ 
SFMC thermal feedback -- supernovae explosion   & (Section \ref{method-sfmc}) &$\,\,\,\,\,\,\,\,\,$\textcircled{}&$\,\,\,\,\,$\textcircled{}&$\,\,\,\,\,$\textcircled{}\\    
MBH accretion\tnote{\textdaggerdbl}  & (Section \ref{method-mbh}) &$\,\,\,\,\,\,\,\,\,$\textcircled{}&$\,\,\,\,\,$\textcircled{}&$\,\,\,\,\,$$\times$\\
MBH radiative feedback -- ionizing radiation  & (Section \ref{method-mbh}) &$\,\,\,\,\,\,\,\,\,$\textcircled{}&$\,\,\,\,\,$\textcircled{}&$\,\,\,\,\,$$\times$\\  
MBH mechanical feedback -- bipolar winds  & (Section \ref{method-mbh}) &$\,\,\,\,\,\,\,\,\,$\textcircled{}&$\,\,\,\,\,$\textcircled{}&$\,\,\,\,\,$$\times$\\  [1ex]
\hline
\end{tabular} 
\vspace{2pt}
\begin{tablenotes}
\item[\textdagger] {\scriptsize For detailed descriptions on included physics, see referenced sections. $\textcircled{} =$ included, $\times = $ not included.}
\item[\textdaggerdbl] {\scriptsize SFMC $=$ star-forming molecular cloud, MBH $=$ massive black hole.}
\end{tablenotes}
\label{table:suite}  
\end{threeparttable}
\vspace{12pt}
\end{table*}

In addition to the radiation channel, a MBH particle can launch bipolar winds by injecting mass and momentum in its vicinity.
The kinetic power of these subrelativistic winds, as we introduce in the cells that are $2\, \Delta x_{14}$ away from the MBH (see Figure 2 of \citealt{2011ApJ...738...54K} for a schematic description), depends on the accretion rate as 
\begin{align}
P_{\rm w} = 0.5\,\epsilon_{\rm w} L_{\rm BH}(t) = {1\over 2} \dot M_{\rm w} v_{\rm w}^2 
\label{eq:wind}
\end{align}
where $\epsilon_{\rm w}$ encapsulates how much of the mechanical energy of a MBH is turned into winds at the scale we introduce them in the simulation ($\sim 2\, \Delta x_{14}$), and $\dot M_{\rm w}$ is the mass ejection rate of the winds.  
With conservative choices of $\epsilon_{\rm w} = 10^{-4}$ and $\eta_{\rm w} = {\dot M_{\rm w} / \dot M_{\rm BH}} = 0.1$ \citep[based on the prior investigation in \citealt{2011ApJ...738...54K} and an 1-dimensional study in e.g.,][]{2009ApJ...699...89C}, the wind velocity $v_{\rm w}$ when introduced in the simulation is determined by
\begin{align}
v_{\rm w} = c \left( { \epsilon_{\rm w} \epsilon_{\rm r} \over \eta_{\rm w} } \right)^{1/2}  = 3000\,\, {\rm km\, s^{-1}}
\end{align}
for $\epsilon_{\rm r} = 0.1$.  
We intermittently inject the ejecta in a form of collimated bipolar winds of width $5\, \Delta x_{14}$ (Figure 2 of \citealt{2011ApJ...738...54K}) every time the accumulated ejecta mass, $\Sigma \,\dot M_{\rm w} dt$, exceeds a $300\,\, {\rm M}_{\odot}$ threshold. 
The direction of the winds is parallel and anti-parallel to the total angular momentum of the accreted gas up to that point, with an added leeway angle of $<10^{\circ}$.\footnote{The random leeway angle of $<10^{\circ}$, which was absent in \cite{2011ApJ...738...54K}, is introduced to broaden the area affected by the MBH winds.}
The velocities of the surrounding cells are then found by averaging the momenta over the injected wind masses and the preexisting cell masses.  
As in the ejecta from SFMCs, 2\% of the MBH wind mass is considered as metals, as we hope to account for unresolved star formation below resolution occurring in the cell hosting the MBH.

\begin{figure*}
\begin{center}
\vspace{2pt}
\includegraphics[width=0.68\textwidth]{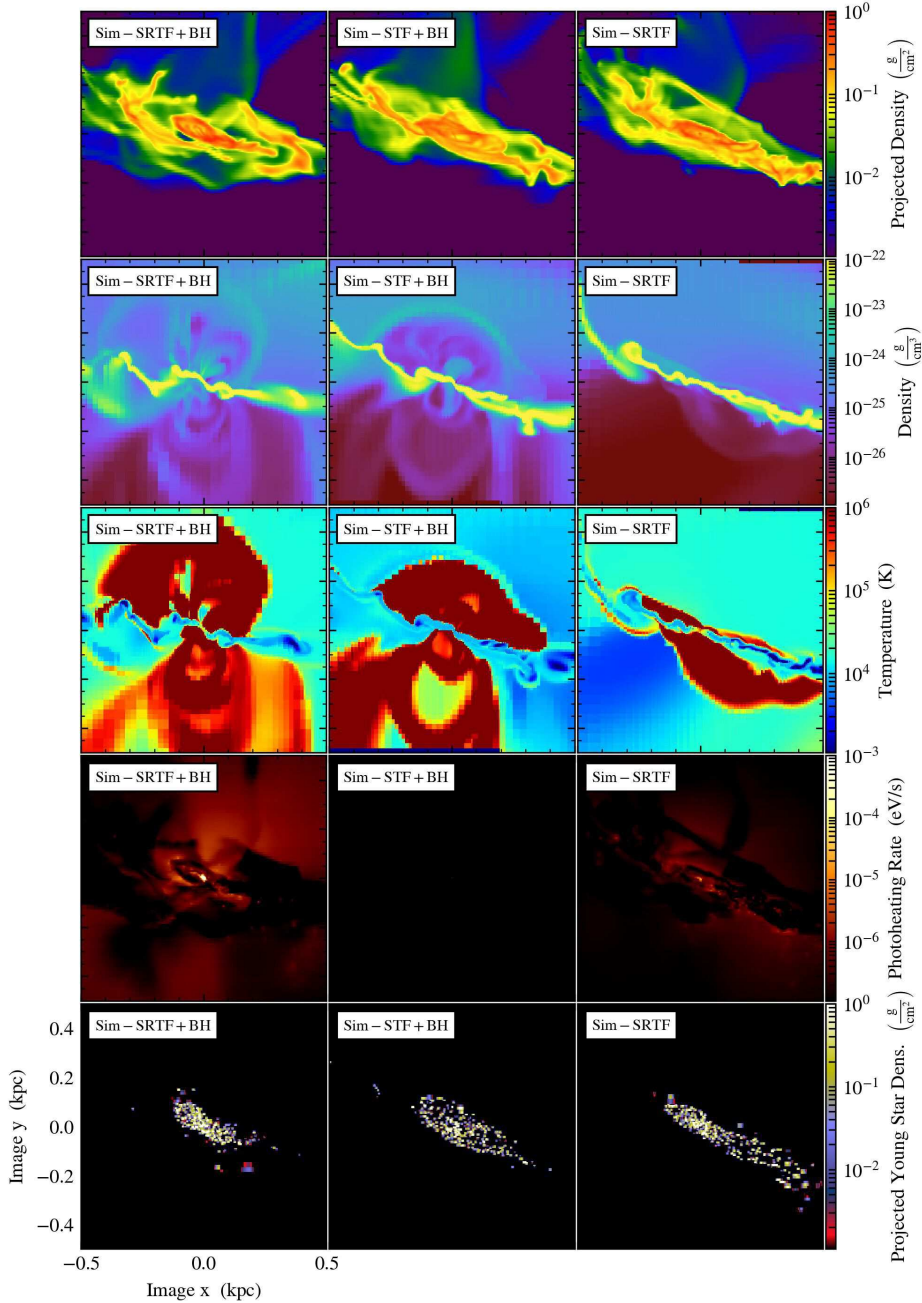}
     \caption{Overview of our simulation suite for a quasar host galaxy with varying input physics, at the end of the run at $z=7.3$  in a $(1\,\,{\rm kpc})^2$ box. Shown here are ({\it from top to bottom row})  gas surface density, sliced density, sliced temperature, projected photoheating rate (density-weighted), and young stellar surface density (age $<3$  Myr),  for the runs with thermal/radiative SFMC feedback and mechanical/radiative MBH feedback ({\it left column; ``Sim-SRTF+BH''}), similar but without radiative SFMC feedback feedback ({\it center column; ``Sim-STF+BH''}), and without a MBH  ({\it right column; ``Sim-SRTF''}).  See Table \ref{table:suite} and Section \ref{method-IC} for more information about the simulation suite.  The line of sight in each panel is perpendicular to the angular momentum of the galactic disk (see also Figure \ref{fig:overview_large}).  Bipolar bubbles created by MBH winds in {\it Sim-SRTF+BH}  and {\it Sim-STF+BH} are prominent in sliced density and temperature ({\it 2nd} and {\it 3rd row}).  Radiation from SFMCs and/or the MBH helps to heat up the gas in {\it Sim-SRTF+BH}  and {\it Sim-SRTF} while such heating is negligible in {\it Sim-STF+BH}  ({\it 4th row}). 
\label{fig:overview_perp}}
\end{center}
\vspace{2pt}
\end{figure*}

\subsection{Initial Condition and Simulation Suite}\label{method-IC}

Since quasars are very rare at $z\gtrsim7$, a large simulation box is necessary. 
Meanwhile, numerical resolution of $<\,$10 pc is essential to resolve the characteristic SFMC scale and the gas flow around a $\gtrsim10^6\,\,{\rm M}_{\odot}$ MBH. 
This high dynamic range is achieved by the ``zoom-in'' initial condition generator {\sc Music}  that uses an adaptive multi-grid Poisson solver \citep{2011MNRAS.415.2101H}, and by the adaptive mesh refinement code {\sc Enzo}.
In particular, we use a set of {\sc Music} parameters that describes a halo that eventually grows into $\sim 10^{13} \,\,{\rm M}_{\odot}$ group at $z=0$ with a relatively quiescent merger history.
This is one of the publicly available cosmological initial conditions identified in a $(60 \,\,\,{\rm comoving} \,\,h^{-1}\,{\rm Mpc})^3$ box by the {\it AGORA} High-resolution Galaxy Simulations Comparison Project  \citep[i.e., initial condition tagged ``1e13q'';][]{2014ApJS..210...14K_short, 2016ApJ...833..202K_short}.\footnote{The website is http://www.AGORAsimulations.org/.  This particular IC, ``1e13q'', has been used in several studies such as \cite{2016ApJ...824..144F}.}  
Here, flat $\Lambda$CDM cosmology consistent with {\it WMAP7+SNe+BAO} is assumed: $\Omega_{\rm m}=0.272$, $\Omega_\Lambda=0.728$, $\sigma_8=0.807$, $n_{\rm s}=0.961$, and $H_0= 70.2\  {\rm km\ } {\rm s}^{-1} {\rm Mpc}^{-1}$ \citep[][see Section \ref{method-hydro}]{2011ApJS..192...18K_short}.
The primary progenitor of our target halo at $z\sim7.5$ is determined as a halo of $M_{\rm vir} \simeq 7\times 10^{10}\,\,{\rm M}_{\odot}$ and $M_{\star} \simeq 8\times 10^{9}\,\,{\rm M}_{\odot}$ with a virial radius $R_{\rm vir} = M_{\rm vir}^{1/3} (H^2 \Omega_{\rm m} \Delta_{\rm c} / 2 G)^{-{1/3}} \simeq 15 \,\,\, {\rm kpc}$ (with $\Delta_{\rm c}=200$).\footnote{The high stellar mass of our target halo at $z\sim 7.5$ is due to an efficient star formation prescription adopted during the lower resolution phase when making the target galaxy (next paragraph).  As will be discussed in Section \ref{stability}, this stellar mass is roughly around the critical value above which MBHs start to grow more rapidly \citep[e.g.,][]{2019arXiv190408431D}.}   
With a $128^3$ root grid and a series of five nested child grids of twice finer resolution each, the equivalent unigrid resolution at level $l=5$ is $4096^3$ (Figure \ref{fig:overview}).  

The simulation is first run with a maximal resolution of 163.0 comoving pc from $z = 100$ to 7.8 (i.e., maximum refinement level $l=12$ with $\Delta \,x_{12} \, = 18.5 \,\,{\rm pc} \,\,\,{\rm at}\,\,\, z=7.8$).
Then, in the next $\sim$ 40 Myrs from $z=7.8$ to 7.5, we simulate the target halo with additional two levels of refinement ($\Delta \,x_{14} \, = 4.79 \,\,{\rm pc} \,\,\,{\rm at}\,\,\, z=7.5$) with the refinement strategy and SFMC physics described in Section \ref{method-hydro} and \ref{method-sfmc}, respectively, but without an accreting MBH. 
This run sets up a relaxed, well-resolved high-$z$ galaxy at $z=7.5$ with which we vary physics to build a suite of simulations (Figures \ref{fig:overview} and \ref{fig:overview_large}). 
Then finally at  $z=7.5$, we restart the calculation with a $1.8\times10^6\,\,{\rm M}_{\odot}$ MBH implanted at the target galaxy's center, while turning on the MBH physics discussed in Section \ref{method-mbh} to run for another $\sim 25$ Myrs until $z=7.3$ ({\it ``Sim-SRTF+BH''}). 
For comparison, another simulation  is run with similar physics inputs but without the {\it radiative} channel of SFMC feedback ({\it ``Sim-STF+BH''}).  
We also have a control run with all the SFMC physics but without a MBH ({\it ``Sim-SRTF''}). 
Table \ref{table:suite} summarizes the suite of simulations we perform.  
Note that the production simulations' run time, $\sim 25$ Myrs, is of the order of the Salpeter timescale, $\tau_{\rm Salpeter}$, a characteristic timescale for a black hole growth at the Eddington accretion limit.

Two points about our experiment are noted:  
{\it (a)} By applying high-resolution refinement only well into the galaxy's evolution, we save the computational expense to simulate the galaxy for a galactic dynamical time, but focus instead on the interaction of a MBH and its neighborhood for a typical lifetime of SFMCs or a rotational timecale of the galactic core region.
{\it (b)} Thus, given a relatively short simulation time, we choose a slightly higher MBH seed mass for the halo mass, in an attempt to observe substantive effect by the MBH in the galactic inner region --- a reasoning similar to \cite{2017MNRAS.465.3291W}. 
Our MBH seed mass of $1.8\times10^6\,\,{\rm M}_{\odot}$ assumes a sizable prior buildup from $\sim 10^2\,\,{\rm M}_{\odot}$ to $\sim 10^6\,\,{\rm M}_{\odot}$ before $z=7.5$. 
For this reason, the reported results should not be regarded as a general picture of MBHs at this redshift (for more discussion on the seed mass, see Section \ref{method-mbh}).

\hspace{5mm}

\begin{figure*}
\begin{center}
\includegraphics[width=0.9\textwidth]{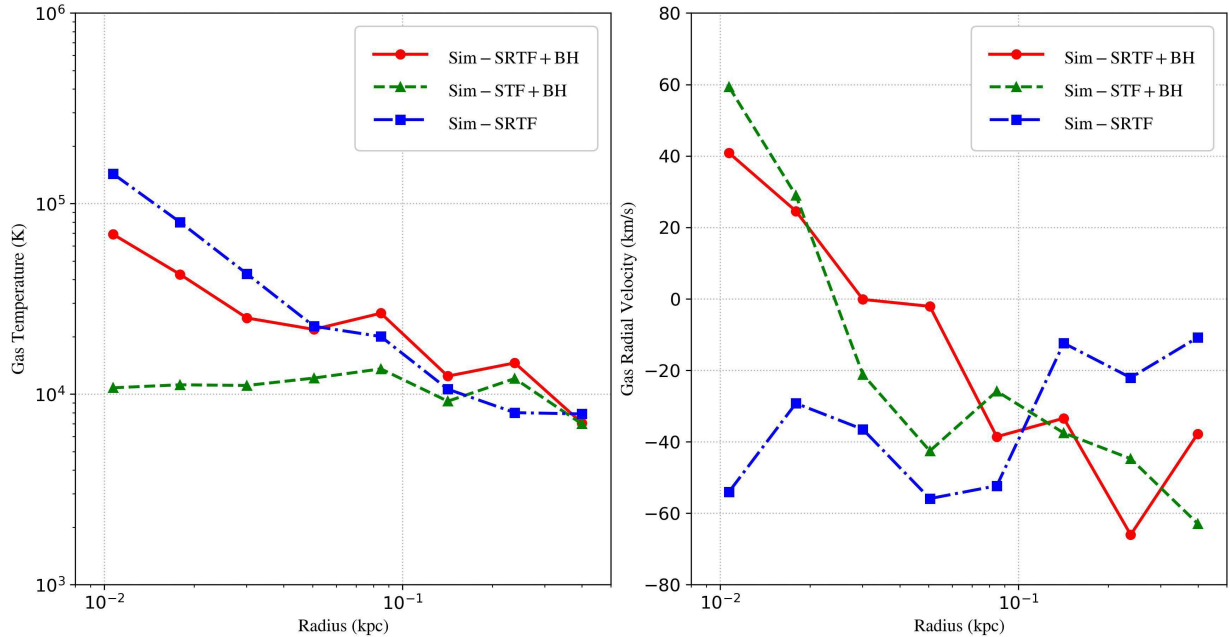}
\vspace{4pt}
     \caption{Mass-weighted radial profiles of temperature ({\it left}) and radial velocity ({\it right}) centered on the MBH at $z=7.3$.  The red solid line, green dashed line, and blue dot-dashed line represent the runs with thermal/radiative SFMC feedback and mechanical/radiative MBH feedback ({\it ``Sim-SRTF+BH''}), similar but without radiative SFMC feedback feedback ({\it ``Sim-STF+BH''}), and without a MBH  ({\it ``Sim-SRTF''}).  See Table \ref{table:suite} and Section \ref{method-IC} for more information about the simulation suite.  The radiation from SFMCs is responsible for the higher temperature of gas within $<100$ pc from the galactic center in {\it Sim-SRTF+BH} and {\it Sim-SRTF} ({\it left}).  The wind feedback by the MBH blows the gas away from the center in {\it Sim-SRTF+BH} and {\it Sim-STF+BH}, a feature absent in {\it Sim-SRTF} ({\it right}). 
\label{fig:profile_gas}}
\end{center}
\vspace{4pt}
\end{figure*}

\section{RESULTS}\label{results}

The reported simulations and analyses were performed on various computational resources including the {\it Happiness} cluster at Seoul National University and the {\it Pleiades} cluster at the NASA Ames Research Center, among others.  
Towards the end of the simulation at $z = 7.3$, each simulation snapshot typically contains a total of $\sim 2\times10^8$ computational elements: $\sim 1.2\times10^8$ particles (dark matter, SFMC, MBH) and $\sim 9\times10^7$ gas cells in $\sim 4\times 10^3$ grid patches.  
With measures to reduce the number of radiation sources (see footnote \ref{footnote:rt}), the radiation transfer module in our code typically handles $\lesssim 3\times 10^4$ ray-emitting sources in {\it Sim-SRTF+BH} and {\it Sim-SRTF}.  
We now analyze the three simulations listed in Table \ref{table:suite} to examine the interplay between the newly introduced SFMC and MBH physics.  
In all subsequent analyses we utilize a code-independent analysis platform {\tt yt} \citep{yt}.\footnote{The website is http://www.yt-project.org/.}

\hspace{1mm}

\subsection{Stellar and MBH Feedback Suppress Runaway Star Formation Locally}\label{results-sf}

Comparing a suite of quasar-host simulations carried out with high spatial resolution ($\Delta \,x_{14} \, = 4.79 \,\,{\rm pc} \,\,\,{\rm at}\,\,\, z=7.5$) and high fidelity physics, we are able to investigate how and how much stellar (SFMC) feedback suppresses star formation in the vicinity of a MBH, and change the ISM structure of a quasar-host galaxy.
It will then lead us to a better picture of how MBHs might have accumulated their masses in the high-$z$  Universe (to be discussed in Section \ref{results-mbh}).  

Figure \ref{fig:overview_perp} overviews our target galaxy at $z=7.3$ in various quantities in a $(1 {\,\,\rm kpc})^2$ field of view from the galactic edge-on angle.   
First, sliced density and temperature plots ({\it 2nd} and {\it 3rd rows}) exhibit the impact of bipolar winds by the MBH (mechanical feedback) that blow surrounding gas away  in {\it Sim-SRTF+BH} and {\it Sim-STF+BH}, leading to mushroom-shaped hot bubbles above and below the disk plane.  
We find that the energetic outflows driven by the MBH affect most critically the gas right above and below the disk, but not so much the gas {\it on} the disk.
Meanwhile, the projected photoheating rates in the gas ({\it 4th row}) reveal how many photons have escaped the dense gas layer right next to the photon sources such as SFMCs and MBHs, and then affect the interstellar gas in the galaxy.
UV radiation from SFMCs and/or the MBH helps to heat up the gas in {\it Sim-SRTF+BH}  and {\it Sim-SRTF}, while such heating is negligible in {\it Sim-STF+BH} (note that only the single MBH radiates photons in the latter case).  

Our qualitative observations can be checked in the radial profiles  in Figure \ref{fig:profile_gas} of temperature and radial velocity in the inner core of the galaxy.  
The left panel of Figure \ref{fig:profile_gas} shows the mass-weighted radial profiles of gas temperature. 
Within a $\sim 100$ pc sphere centered on the MBH, the photoheating radiation from (mostly) young SFMCs helps to heat the gas up to $\sim 10^5$ K  in {\it Sim-SRTF+BH} and {\it Sim-STRF} --- by making gas densities lower and cooling times longer, thus the SN feedback more efficient (so the SN ejecta cools more slowly in the presence of a radiation field).
The nearly constant temperature of $\sim 10^{4}$ K for {\it Sim-STF+BH} indicates that the photons from the MBH alone (see Table \ref{table:suite}) are not enough to raise the temperature above $\sim 10^{4}$ K, or fails to escape the thick layers of neutral hydrogen near the MBH (to be discussed again in Sections \ref{results-mbh} and \ref{fesc}). 
It also suggests that the high temperatures in the galactic inner region in {\it Sim-SRTF+BH} and {\it Sim-STRF} are due in most part to radiation from SFMCs (see Section \ref{fesc}).  
Then the right panel of Figure \ref{fig:profile_gas} shows a radial velocity profile evaluated from the location of the MBH, and how the gas dynamics in the galactic core region is changed by the MBH mechanical feedback.
At large scales, there generally exist prominent gas inflows for a young, fast-growing galaxy like the one presented here, as are seen in all three simulations at outer radii ($\gtrsim 50$ pc). 
However, the MBH winds blow the gas away in the direction perpendicular to the disk plane in {\it Sim-SRTF+BH} and {\it Sim-STF+BH}, making the gas in the inner layer ($\lesssim 50$ pc) move outward {\it on average}.
It is despite the fact that the inflow on the galactic disk plane is largely unaffected by the MBH feedback due to its thickness. 
Clearly, this is a feature absent in a run without the MBH feedback, {\it Sim-SRTF}.\footnote{Note that because radiation pressure is not considered in the reported simulations, MBH's radiation alone would not drive outflows  in {\it Sim-SRTF+BH} or {\it Sim-STF+BH}.  Had we included the radiation pressure by the MBH --- and SFMCs ---, the gas outflows might have been enhanced.}  

\begin{figure*}
\begin{center}
\includegraphics[width=0.98\textwidth]{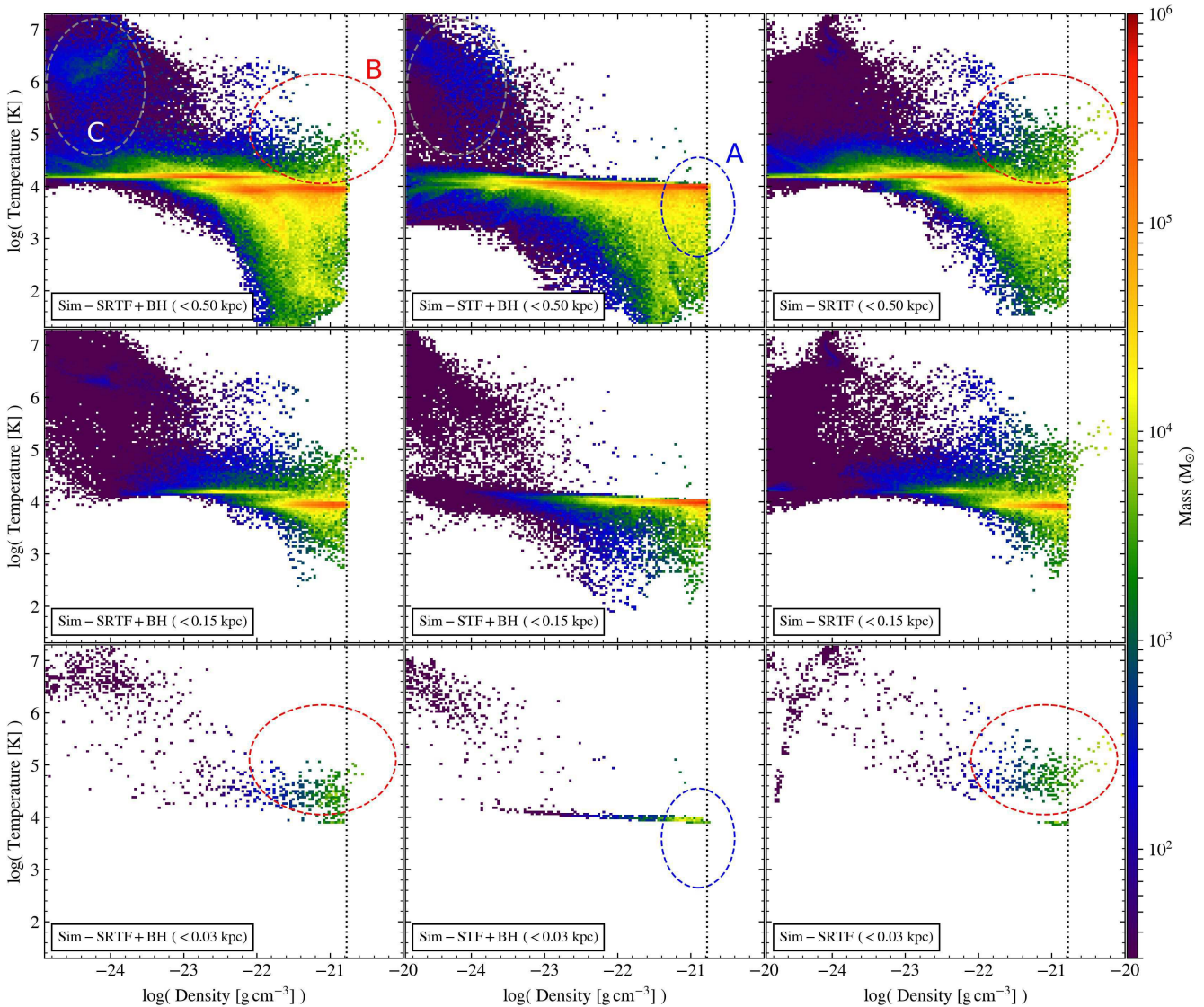}
\vspace{4pt}
     \caption{Density-temperature joint probability distribution functions (PDFs) in spheres of successively smaller radii ({\it from top to bottom row}) centered on the MBH at $z=7.3$,  for the runs with thermal/radiative SFMC feedback and mechanical/radiative MBH feedback ({\it left column; ``Sim-SRTF+BH''}), similar but without radiative SFMC feedback feedback ({\it center column; ``Sim-STF+BH''}), and without a MBH  ({\it right column; ``Sim-SRTF''}).  See Table \ref{table:suite} and Sectio7 \ref{method-IC} for more information about the simulation suite.  Colors represent the gas mass in each 2-dimensional bin.  The star (SFMC) formation threshold density is denoted by a black dotted line in each panel ($n_{\rm thres}$; see Section \ref{method-sfmc}).  
     Without UV radiation from SFMCs, in {\it Sim-STF+BH}, cold dense gas turns into SFMCs at $n_{\rm thres}$ ({\it region ``A''}).  In contrast, UV radiation from SFMCs  and/or from the MBH helps to heat up gas close to the MBH to $\sim10^5$ K in {\it Sim-SRTF+BH}  and {\it Sim-SRTF} making the gas stable against fragmentation ({\it region ``B''}). Note also that the bipolar wind feedback from the MBH enhances and maintains the hot diffuse gas of $\gtrsim 10^6\,\,{\rm K}$ in {\it Sim-SRTF+BH} and {\it Sim-STF+BH} ({\it region ``C''}).
\label{fig:overview_pdf}}
\end{center}
\vspace{8pt}
\end{figure*}

Figure \ref{fig:overview_pdf} shows in detail the structure of the ISM affected by stellar (SFMC) and MBH feedback within different enclosing radii.
Each panel shows a density-temperature joint probability distribution function (PDF) with a black, vertical dotted line denoting the density threshold for SFMC formation, $n_{\text{thres}}$ (see Section \ref{method-sfmc}).
Without UV radiation from SFMCs, in {\it Sim-STF+BH}, cold dense gas instantly turns into SFMCs at $n_{\rm thres}$ leaving no gas over the threshold  ({\it region ``A''}).  
In contrast, the UV radiation from SFMCs  and/or from the MBH helps to heat up the dense gas close to the MBH to $\sim10^5$ K in {\it Sim-SRTF+BH}  and {\it Sim-SRTF} making the gas stable against fragmentation.  
Some hot, dense gas cells irradiated by SFMCs or the MBH still exist beyond the threshold as they do not turn into stars ($t_{\rm{cool}} > t_{\rm{dyn}}$; {\it region ``B''}).  
The hot diffuse gas with $T \gtrsim 10^6\,\,{\rm K}$ and $\rho \lesssim 10^{-23}\,\,{\rm g\,cm^{-3}}$ formed via supernova feedback is visible in all three simulations, but is particularly enhanced by the wind feedback from the MBH in {\it Sim-SRTF+BH} and {\it Sim-STF+BH} ({\it region ``C''}).
     
\begin{figure*}
\begin{center}
\includegraphics[width=0.9\textwidth]{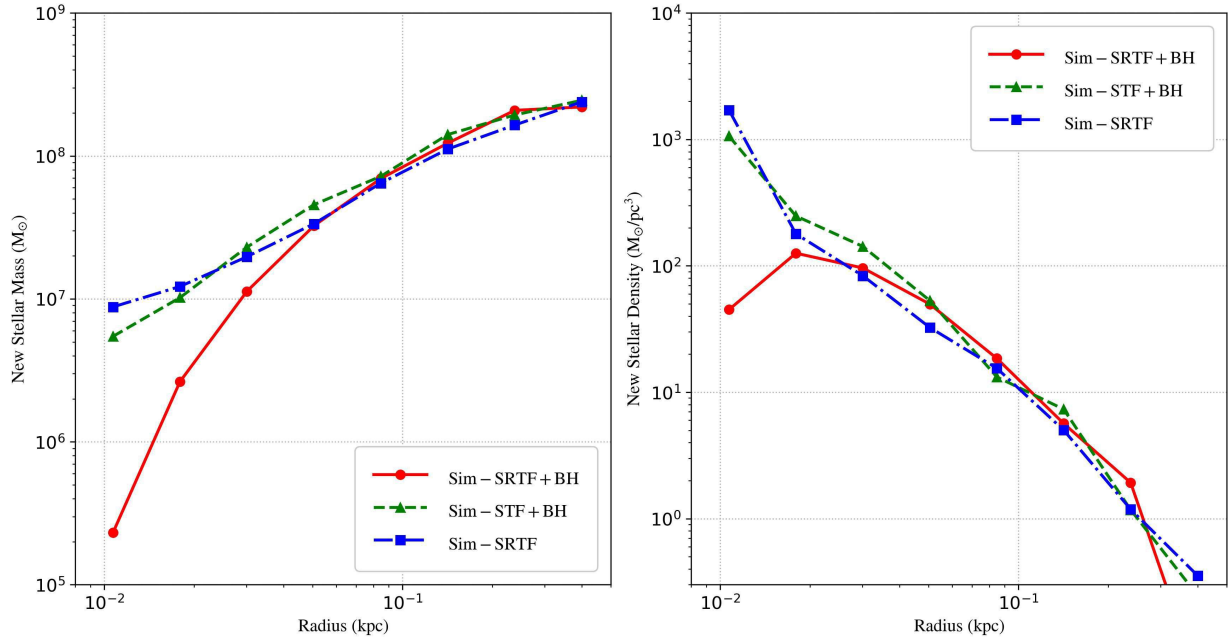}
\vspace{4pt}
     \caption{Enclosed mass ({\it left}) and radially-averaged density profiles ({\it right}) of ``new'' SFMCs (those who were created between $z=7.5$ and 7.3 after we vary the SFMC/MBH physics between different runs) in a 150 pc sphere centered on the MBH at $z=7.3$.  See Table \ref{table:suite} or the caption of Figure \ref{fig:profile_gas} for more information about the simulation suite.  Locally suppressed star formation in {\it Sim-SRTF+BH} results in a considerable reduction of new stellar mass in the galactic core region.  Considering the findings in Figures \ref{fig:overview_perp} to \ref{fig:overview_pdf}, note that only the combination of stellar (SFMC)$+$MBH radiation  and MBH wind feedback leads to such suppression.   
\label{fig:profile_new_star}}
\end{center}
\end{figure*}

As Figures \ref{fig:overview_perp} to \ref{fig:overview_pdf} have illustrated, the hot temperature induced by the SFMC and MBH radiation, and the diverging flow of gas by the MBH-driven winds suppress the formation of SFMCs locally in {\it Sim-SRTF+BH}.
The gas cells in the galactic center have harder time to fulfill all the SFMC formation conditions, such as a short cooling time ($t_{\rm{cool}} < t_{\rm{dyn}}$) and a converging velocity flow ($\nabla \cdot v < 0$). 
Figure \ref{fig:profile_new_star} shows the resulting distribution of new SFMCs (stellar mass) in the galaxy's inner region.
The red line depicts the suppressed star formation in {\it Sim-SRTF+BH} between $z=7.5$ and 7.3 resulting in a considerable reduction of ``new'' stellar masses (SFMCs that were created between $z=7.5$ and 7.3 after we vary the SFMC/MBH physics between different runs), in enclosed mass ({\it left panel}) and radially-averaged density ({\it right panel}) profiles.   
As our previous findings suggest, note that only the combination of stellar (SFMC)$+$MBH radiation and MBH wind feedback leads to such suppression.   

We caution that because the run times of the simulations reported here are relatively short owing to various numerical limitations (e.g., $\sim 25$ Myrs for {\it ``Sim-SRTF+BH''}), we have observed only the beginning of how SFMC and MBH feedback will transform the galaxy as a whole.  
An attempt to acquire a more complete picture of the galactic transformation --- by advancing the simulation for a longer time with better optimizations --- is currently being made (see Section \ref{future}).

\hspace{1mm}

\subsection{Stellar and MBH Feedback Help The MBH Grow \\
By Retaining The Fuel For Accretion}\label{results-mbh}

\begin{figure*}
\begin{center}
\vspace{4pt}
\includegraphics[width=0.9\textwidth]{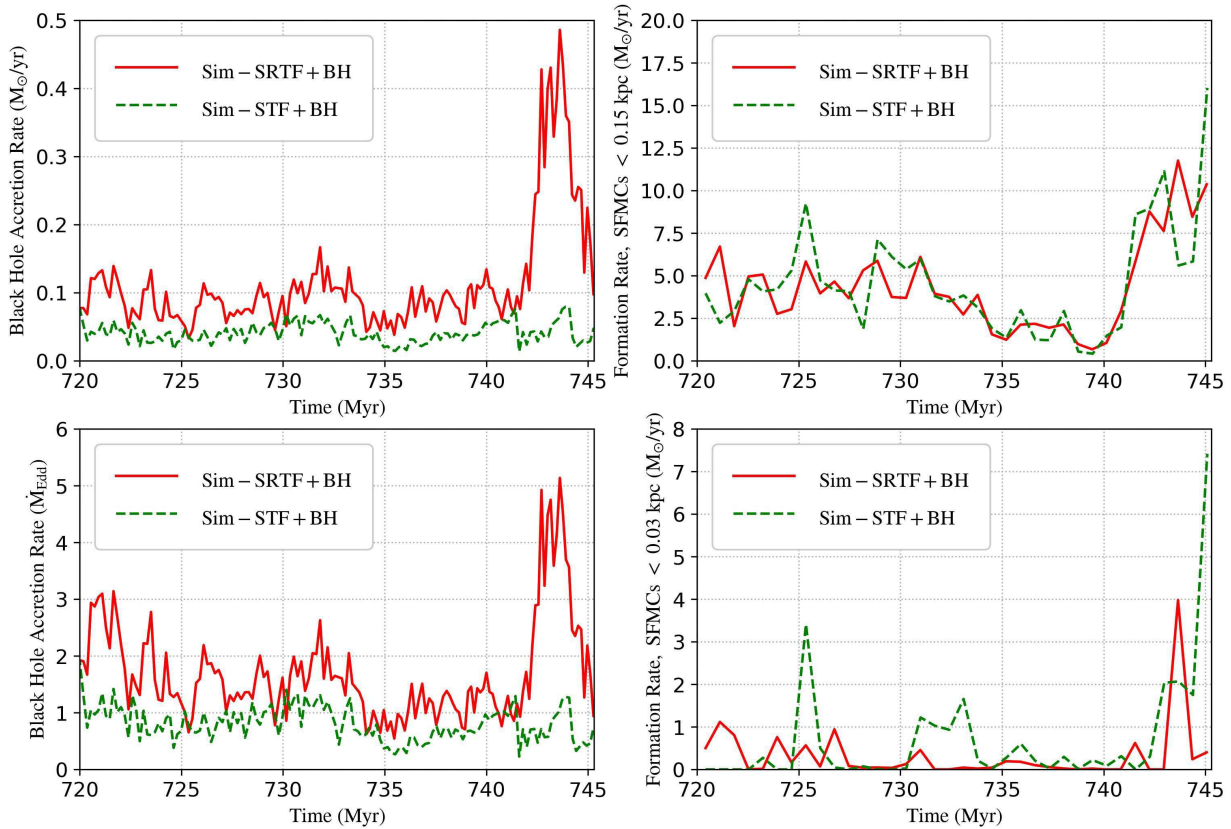}
\vspace{4pt}
     \caption{MBH accretion rate in the unit of ${\rm M}_{\odot} {\rm yr}^{-1}$ ({\it top left}) and $\dot{M}_{\rm Edd}$ ({\it bottom left}), and the formation rates of stars (SFMCs) that are located within 150 pc from the MBH at $z=7.3$ ({\it top right}) and within 30 pc  ({\it bottom right}).  Including the radiation from SFMCs ({\it Sim-SRTF+BH}) suppresses star formation in the galactic core region ({\it bottom right;} see also Figure \ref{fig:profile_new_star}) and helps to retain gas that eventually falls in to the MBH ({\it top left}).
\label{fig:bhar}}
\end{center}
\vspace{8pt}
\end{figure*}

\begin{figure*}
\begin{center}
\includegraphics[width=0.67\textwidth]{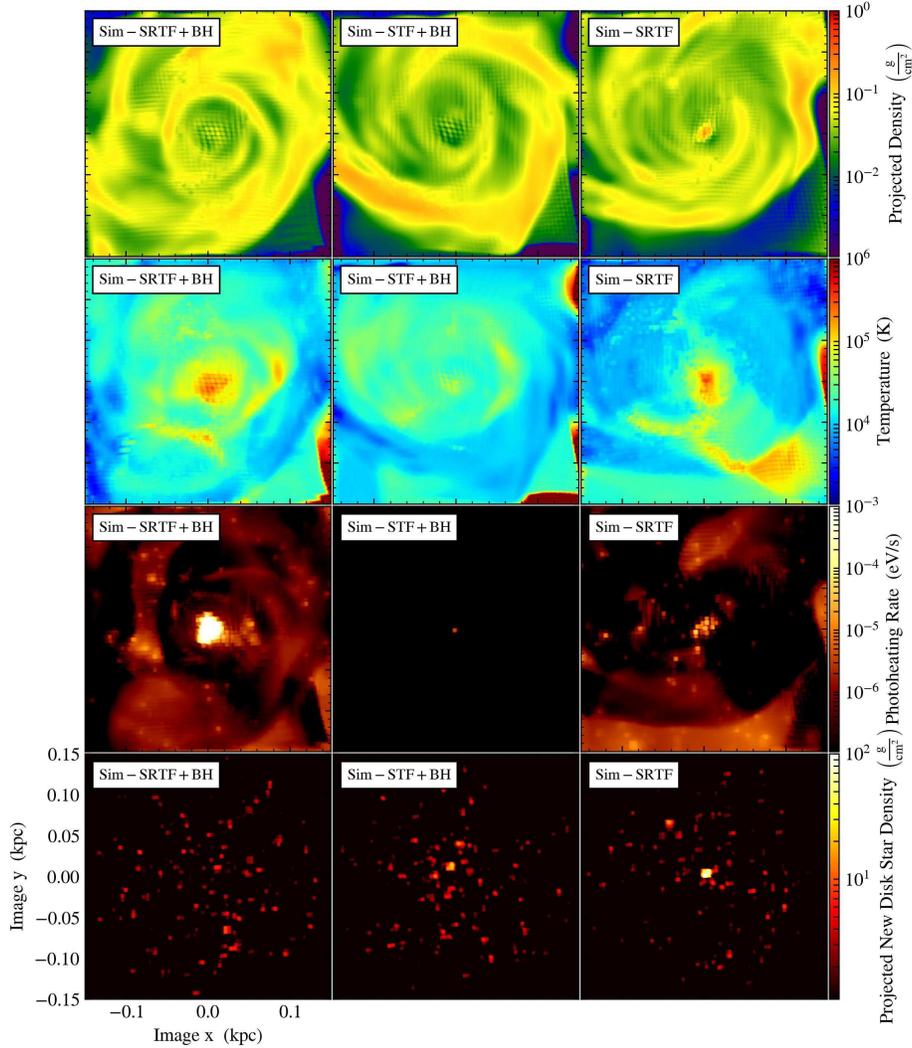}
\vspace{5pt}
     \caption{Similar to Figure \ref{fig:overview_perp} but in a smaller $(300\,\,{\rm pc})^2$ box centered on the MBH at $z=7.3$.  Shown here are  ({\it from top to bottom row}) gas surface density, projected temperature (density-weighted), projected photoheating rate (density-weighted), and surface density of ``new'' disk stars (SFMCs that are within $<$ 30 pc above and below from the disk plane and are created after $z=7.5$ when we variate the SFMC/MBH physics in Sections \ref{method-sfmc} and \ref{method-mbh} between different runs). For more information about the three different simulation runs ({\it from left to right column}), see Table \ref{table:suite} or the caption of Figure \ref{fig:profile_gas}.  The line of sight in each panel is parallel to the angular momentum of the inner galactic disk.  Radiation from SFMCs  and/or the MBH helps to keep hot gas in {\it Sim-SRTF+BH}  and {\it Sim-SRTF} while such heating is negligible in {\it Sim-STF+BH}  ({\it 2nd} and {\it 3rd row}).  In particular, when all feedback channels from both SFMCs and the MBH are included, SFMC formation is suppressed in the close vicinity of the MBH ({\it 4th row}).
\label{fig:overview_para}}
\end{center}
\vspace{6pt}
\end{figure*}

Now we specifically focus on how the MBH acquires its mass, and how SFMC and MBH feedback have affected its accretion process in our suite of simulations (Table \ref{table:suite}).   
Figure \ref{fig:bhar} displays the gas accretion rate on to the central MBH ({\it left panels}) and the formation rates of stars (SFMCs) that are located within 150 pc from the MBH at $z=7.3$ ({\it top right panel}), and within 30 pc  ({\it bottom right panel}).
Combined with our findings in Figure \ref{fig:profile_new_star}, Figure \ref{fig:bhar} implies that the SFMC formation in {\it Sim-SRTF+BH} is suppressed mostly only within the galaxy's inner core region ($\lesssim 50$ pc). 
The unused interstellar gas remains near the MBH which could otherwise have been rapidly consumed through runaway star formation (see also Figures \ref{fig:overview_pdf} and \ref{fig:overview_para}).  
In the meantime, MBH accretion rate of {\it Sim-SRTF+BH} is continuously higher than that of {\it Sim-STF+BH}.\footnote{The elevated MBH accretion rate star formation rate at 740-745 Myr  is thought to be related to a recent close encounter with a smaller galaxy.  We hope to investigate this merging event in future research (see Section \ref{future}).}
Indeed, the MBH  in {\it Sim-SRTF+BH} after just $\sim$ 25 Myr grew by $ 2.9\times10^6\,\,{\rm M}_{\odot}$ with a growth rate of $0.113\,\, {\rm M}_{\odot} {\rm yr}^{-1}$.
This is approximately 1$-$3 times the Eddington rate for the MBH mass of $\sim 2\times10^6\,\,{\rm M}_{\odot}$ ({\it bottom left panel} of Figure \ref{fig:bhar}).  
Note that the Eddington limit is not imposed in our accretion estimation (see the related discussion in Section \ref{method-mbh}). % Eddington rate 0.02 Ms/yr * (M_BH / 1e6 Ms)  
By contrast, the MBH mass in {\it Sim-STF+BH} was increased by $1.1\times10^6\,\,{\rm M}_{\odot}$ at the rate of $0.041\,\, {\rm M}_{\odot} {\rm yr}^{-1}$.  
These observations strongly indicate that the unconsumed gas  in {\it Sim-SRTF+BH} fuels the enhanced growth of the MBH when compared with {\it Sim-STF+BH}.

\begin{figure*}
\begin{center}
\includegraphics[width=0.9\textwidth]{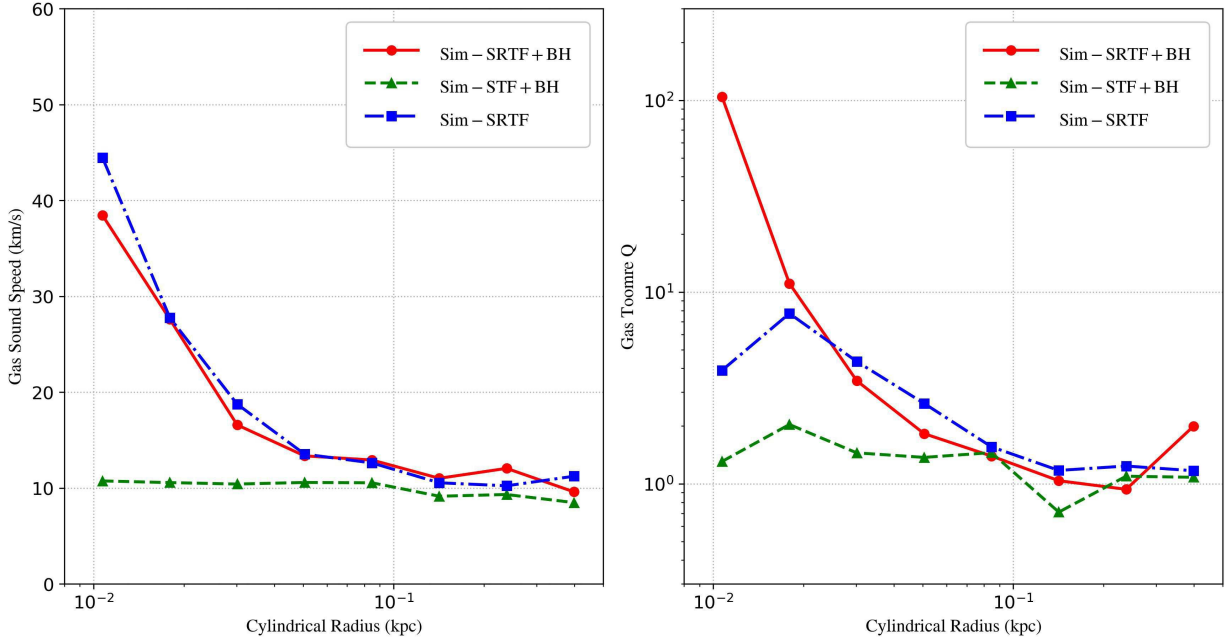}
\vspace{4pt}
     \caption{Mass-weighted cylindrical profiles of the sound speed ({\it left}) and Toomre Q parameter of gas ({\it right}) in a disk centered on the MBH at $z=7.3$.  The disk has a thickness 60 pc.  Due to the radiation mostly by SFMCs, the gas at the galactic core region in {\it Sim-SRTF+BH} and {\it Sim-SRTF} has large sound speeds ({\it left}).  The gas Toomre Q parameter indicates that the gas within a cylindrical radius of $\sim100$ pc from the MBH in {\it Sim-SRTF+BH} is mostly stable against fragmentation, while the gas in {\it Sim-STF+BH} is marginally unstable ({\it right}).
\label{fig:profile_Toomre}}
\end{center}
\vspace{5pt}
\end{figure*}

To take a deeper look into the neighborhood around the accreting MBH, in Figure \ref{fig:overview_para} we display the snapshots of the target galaxy in boxes of $(300\,\,{\rm pc})^3$ centered on the MBH at $z=7.3$ from the face-on angle of its disk plane. 
In the analyses hereafter (for Figures \ref{fig:overview_para}-\ref{fig:profile_Toomre}), the disk is defined as a plane perpendicular to the angular momentum vector of the gas within a 150 pc radius.  
As seen in Figure \ref{fig:profile_gas}, a core of hot gas above $\gtrsim 10^5 {\rm K}$ occupies the galaxy's center in {\it Sim-SRTF+BH} and {\it Sim-SRTF} ({\it 2nd row}). 
This hot core is created by radiation from mostly SFMCs, as discussed in Section \ref{results-sf}, while such photoheating effect is nearly absent in {\it Sim-STF+BH}  ({\it 3rd row}; photons from the MBH rarely escapes the galaxy's center most of the time in {\it Sim-STF+BH} due to thick layers of neutral hydrogen). 
And the ``new'' stellar density in the disk of {\it Sim-SRTF+BH} shows a sign of locally-suppressed star formation in the core region when compared with {\it Sim-STF+BH} and {\it Sim-SRTF} ({\it 4th row};  for the definition of ``new'' stellar mass, see the caption of Figure \ref{fig:overview_para}).
As discussed in Section \ref{results-sf} and Figure \ref{fig:profile_new_star}, only the combination of stellar (SFMC)$+$MBH radiation and MBH winds leads to such suppression. 

The relation between the SFMC$+$MBH feedback and the state of the gas around the MBH can be nicely explained by the cylindrical profiles of the sound speed (measure of gas pressure) and the Toomre Q parameter (measure of gas stability) in Figure \ref{fig:profile_Toomre}.
On the left panel, the high value of gas sound speed, $c_{\rm s} \sim (k_{\rm B} T/ m_{\rm p})^{1/2}$, in {\it Sim-SRTF+BH} and {\it Sim-SRTF} is a direct consequence of photoheating radiation by SFMCs discussed in Figure \ref{fig:profile_gas}.  
With this $c_{\rm s}$, our slightly modified version of the Toomre Q parameter is displayed on the right panel.\footnote{
We modify the Toomre Q parameter on the gaseous disk at radius $r$ as 
\begin{align}
Q (r) ={{c_{\rm s}\, \Omega(r)} \over { \pi G (\Sigma_{\rm gas} + \Sigma_{\rm star})  } }  \label{eq:Toomre}
% {{c_{\rm s} \kappa }\over { \pi G \Sigma_{\rm gas} } }
\end{align}
where $\Sigma_{\rm gas}$ and $\Sigma_{\rm star}$ are the gas and stellar surface density, respectively \citep{1964ApJ...139.1217T, 1994ApJ...427..759W}.  
%$\sigma_{\rm gas} = (<\!\!\!{\bold v}^2\!\!\!>\! - \!<\!\!\!{\rm v}_{\phi}\!\!\!>^2)^{1/2}$ is the velocity dispersion of the gas 
The usual epicycle frequency $\kappa(r) = {\left[ r{d\Omega^2(r) / dr} + 4 \Omega^2(r) \right]}^{1/2}$ is replaced by the angular speed $\Omega(r)$ as these two values are in a factor of 2 difference \citep{2009ApJ...696...96W}.  
%In the same manner, the Q parameter for the stellar disk is now
%\begin{align}
%Q_{\rm star} = {{\sigma_{\rm star} \Omega } \over { 3.36 G (\Sigma_{\rm gas} + \Sigma_{\rm star}) } }
%{{\sigma_{\rm star} \kappa }\over { 3.36 G \Sigma_{\rm star} } } 
%\end{align}
%given the stellar velocity dispersion $\sigma_{\rm star}$.
%where ${\rm v}_{\phi}$ means the tangential velocity and $<\!\!\cdot\!\!>$ the azimuthal average in each disk annulus.  
}
It shows $Q(r) > 1$ at $r \lesssim 100\,\, {\rm pc}$ for {\it Sim-SRTF+BH} and {\it Sim-SRTF}, implying that the rotating disk in the galactic core is largely stable against fragmentation. 
This strengths our notion that the stellar UV radiation, combined with the MBH feedback, helps to retain the gas that could otherwise have been consumed through fragmentation and ensuing star formation.  
This gas may remain warm on the disk until it eventually accretes on to the MBH.  
On the other hand, the small sound speed in {\it Sim-STF+BH} makes the disk gas  marginally unstable with $Q(r) \sim 1$ ($T\sim 10^4\,\,{\rm K}$ in Figure \ref{fig:profile_gas}), leading to SFMC formation and thus less fuel for the MBH growth.

Our experiments demonstrate that the interplay between gas, stars, and MBHs that were never realized in previous simulations may contain critical information about the expeditious growth of MBHs at high $z$. 
Specifically, our results emphasize the need to properly model the {\it circumnuclear} region in the vicinity of a MBH when performing a high-resolution simulation to probe the growth of a MBH \citep[for more information about interactions between gas, stars, and MBHs in the circumnuclear region of a galaxy, see e.g.,][]{2012AdAst2012E..15N, 2013ApJ...763...62A, 2013A&A...560A..34W, 2013ApJ...777L..14F, 2015ApJ...803...81N, 2018MNRAS.475.5688B}.
Interested readers are also referred to Section \ref{future} on the need for even higher-resolution simulations.

\vspace{5mm}

\section{DISCUSSION}\label{discussion}

Before proceeding to conclude the article, we draw the readers' attention to two points that are worthy of brief discussions.

\subsection{How to Retain The Fuel To Feed A MBH}\label{stability}

The argument in Section \ref{results} that the various modes of SFMC$+$MBH feedback help to retain the fuel for the MBH growth that could otherwise have been consumed through fragmentation, is reminiscent of other numerical studies testing rapid black hole growth scenarios.
For example, some groups have claimed that the Lyman-Werner background radiation from nearby galaxies could induce halos to collapse isothermally to high densities without fragmenting into smaller clumps.  
By keeping the gas stable against fragmentation, this mechanism may provide an important pathway to a so-called direct collapse black hole of intermediate size, $10^3-10^5\,\,{\rm M}_{\odot}$ \citep[DCBH; e.g.,][]{2015MNRAS.451.2352K, 2017NatAs...1E..75R, 2018NatAs...2..987B, 2019Natur.566...85W}.
Others have asserted that gravo-turbulence in multi-scale gas inflows in gas-rich galaxy mergers could prevent fragmentation from occurring, and put the gas Toomre Q parameter above 1 within 30 pc from the galaxy's center.
It thus presents another way to trigger the emergence of DCBH seeds \citep[e.g.,][]{2015ApJ...810...51M, 2019RPPh...82a6901M}. 
The result of our experiment presented in this paper is broadly in line with these attempts that are trying to find a viable route to direct gas collapse and expedite the growth of MBHs.

Another mechanism that several numerical studies have recently argued to retain a reservoir of gas to fuel a MBH is through building a compact and massive galactic core \citep[e.g.,][and references therein]{2019arXiv190408431D}.  
The growth of a MBH is slow when the supernova feedback pushes the gas out of the galactic potential well, but once above a critical mass of $M_{\star} \sim 10^{10}\,\,{\rm M}_{\odot}$ the MBH grows rapidly since the ``compaction'' near the galaxy center makes the MBH sink to the center and brings the supernova ejecta back to its deep potential \citep[see Figure 8 of][]{2019arXiv190408431D}.  
Our target galaxy used in this paper is just around this critical mass with $M_{\rm vir} \simeq 7\times 10^{10}\,\,{\rm M}_{\odot}$, $M_{\star} \simeq 8\times 10^{9}\,\,{\rm M}_{\odot}$, and the stellar/gas mass within 1 kpc from the center, $M_{\star,\,{\rm <1\,kpc}} \simeq 3\times 10^{9}\,\,{\rm M}_{\odot}$ and $M_{\rm gas,\,<1\,kpc} \simeq 3\times 10^{8}\,\,{\rm M}_{\odot}$, respectively (see Section \ref{method-IC}).   
Indeed, there is a hint that the combination of the compact core (with $\Sigma_{\star, \,{\rm <1\,kpc}} \sim 10^{9}\,\,{\rm M}_{\odot} \,{\rm kpc}^{-2}$) and the SFMC$+$MBH feedback helps retain a reservoir of gas to feed the MBH.   
However, it remains to be seen if this behavior will be sustained beyond the simulation time reported here, $\sim 25$ Myrs.

\vspace{1mm}

\subsection{Galactic Escape Fraction}\label{fesc}

\begin{figure*}
\begin{center}
\includegraphics[width=1.02\linewidth]{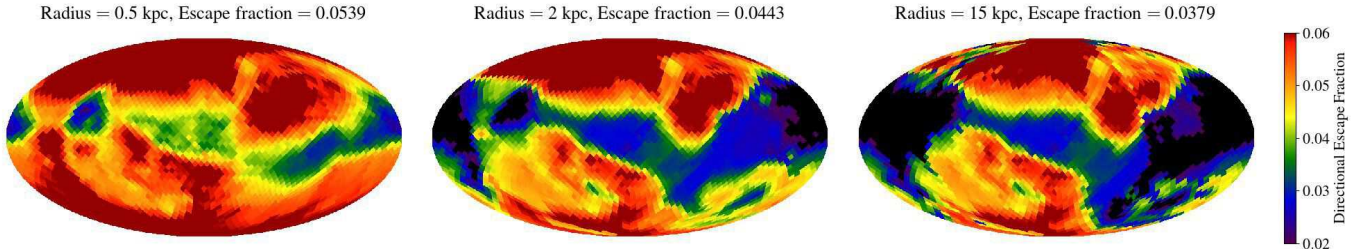}
     \caption{All-sky maps of UV photon escape fractions at 0.5, 2.0 and 15 kpc ({\it from left to right}) along different lines of sight from the SFMCs.  The equator plane of the map is perpendicular to the angular momentum vector of the galactic disk.  The plots are obtained by post-processing the optical depth of hydrogen-ionizing  photons in each direction in the \textit{Sim-SRTF+BH} run at $z=7.3$.  At 15 kpc ($\simeq R_{\rm vir}$; see Section \ref{method-IC}), the galactic {\it average} escape fraction, $f_{\rm{esc}}$, is 0.0379 at this epoch.  $f_{\rm{esc}}$ stays around $\lesssim 0.05$ throughout  the simulation time.  UV photons have harder time to penetrate the thicker neutral hydrogen along the galaxy's disk plane which is shown here as bluer color (lower escape fraction) along the equator.
\label{fig:esc_frac}}
\end{center}
\vspace{8pt}
\end{figure*}

Because galaxies are simulated with ionizing photons from SFMCs and MBHs on the fly, we can study how many photons escape a galaxy using the post-processing machinery developed in \cite{2013ApJ...775..109K}.  
Interested readers are referred to the Section 5.2 of \cite{2013ApJ...775..109K} on how an all-sky escape fraction map is built. 
Figure \ref{fig:esc_frac} presents the all-sky maps of escape fraction at various distances along different lines of sight from the SFMCs in \textit{Sim-SRTF+BH}. 
As noted on top of each panel, the {\it average} escape fraction decreases as the column density of neutral hydrogen increases. 
At 15 kpc ($\simeq R_{\rm vir}$; Section \ref{method-IC}), the {\it average} escape fraction, $f_{\rm{esc}}$, is 0.0379 at this epoch, and it stays around $\lesssim 0.05$ throughout the simulation time.  
UV photons have harder time to penetrate the thicker neutral hydrogen along the galaxy's disk plane, shown here as bluer color (lower escape fraction) along the equator.

Though in Figure \ref{fig:overview_perp} ({\it 4th row}) there seems to be an indication that {\it Sim-SRTF+BH} has more enhanced UV escape fraction than {\it Sim-SRTF} has, we find no such trend in the $\lesssim 20$ snapshots we stored throughout the simulation time.  
In other words, {\it Sim-SRTF+BH} and {\it Sim-SRTF} do not show statistically signifiant difference in $f_{\rm{esc}}$.
We suspect that the $f_{\rm{esc}}$ values of the two runs are similar because the photoionization and clearing of the ISM are done mostly by radiation form SFMCs at these epochs.\footnote{
The total UV luminosity of ``new'' SFMCs in our calculation is
\begin{align}
L_{\rm MC}(t) &= q_{\rm MC}  \, E_{\rm ph}  \, M_{\rm MC}(t)  \nonumber \\
&\sim 6.3\times 10^{46} \times 16.0 \times M_{\rm MC}(t)  \,\,\,{\rm eV\,s^{-1}}   \nonumber \\ 
&\sim 10^{44} \,{\rm erg\,s^{-1}} \left[\,M_{\rm MC}(t) / 10^8 {\rm M}_{\odot}  \right],
\end{align}
from Eq.(\ref{eq:lum_MC}) and Figure \ref{fig:profile_new_star}, whereas the total UV luminosity of a $\sim 2 \times10^6\,\,{\rm M}_{\odot}$ MBH accreting at the Eddington rate  is
\begin{align}
L_{\rm BH,\, UV}(t) &= 0.5\,f_{\rm UV} \epsilon_{\rm r} \dot M_{\rm BH}(t) c^2  \nonumber \\
&\sim 0.5 \times 0.1 \times 0.1 \times \dot M_{\rm BH}(t)  \times c^2    \nonumber \\
&\sim 10^{43} \,{\rm erg\,s^{-1}} [\, \dot M_{\rm BH}(t) / 0.05\,\, {\rm M}_{\odot} {\rm yr}^{-1} ],
\end{align}
 from Eq.(\ref{eq:lum_UV_BH}).  This suggests that for our young, fast-growing target galaxy at $z=7.5$, the UV radiation could be dominated by SFMCs.
}
As seen in {\it Sim-STF+BH} ({\it 4th row} of Figure \ref{fig:overview_perp}), the MBH's UV luminosity alone is not enough to carve out pathways for photons to escape.
The observation that the MBH radiation does not significantly alter $f_{\rm{esc}}$ should however be tested with a run with longer evolution time (see Section \ref{future}).  
AGN contribution to  $f_{\rm{esc}}$ could be little in the particular epoch we tested, but may become important later \cite[for more discussion on the relative contributions to Universe's reionization by quasars and stellar sources, see e.g.,][]{2010Natur.468...49R, 2015ApJ...813L...8M}.

\vspace{5mm}

\section{CONCLUSION}\label{conclusion}

\subsection{Summary}\label{summary}

Using a state-of-the-art simulation technique on an adaptively refined mesh, a suite of high-resolution simulations of a massive $\sim7\times 10^{10}\,\, {\rm M}_{\odot}$ galaxy for $\sim 25$ Myrs at $z\sim7.5$ has been carried out with a $\gtrsim10^6\,\, {\rm M}_{\odot}$ embedded MBH seed, portraying an analogue of a fast-growing MBH seed in a high-$z$ galaxy. 
The high resolution imposed in our simulations allows us to self-consistently incorporate all galactic components and the interactions between them that are important to understand the high-$z$ quasar-host galaxies.  
Our simulations feature, most importantly, radiating stars and MBHs as we explicitly trace their photoionizing radiation through a full 3-dimensional radiative transfer calculation on the fly.  
Additional feedback channels such as supernovae explosions and bipolar winds from a MBH are considered, too  (Sections \ref{method-sfmc} and \ref{method-mbh}). 
To the best of our knowledge, all these sophisticated physics models have rarely been realized with sufficient resolution in a galaxy-scale numerical study regarding high-$z$ MBH-host galaxies.  
In this regard, our approach differs from, and complements, previous studies that have often resorted to thermal or mechanical feedback alone with ad hoc conversion efficiencies, ignoring the coupling processes of SFMC or MBH feedback with the ISM below resolution. 

In this first report, we have focused on the evolution of the MBH seed as well as the inner region of its host galaxy.  
We find that feedback from SFMCs and an accreting MBH helps to suppress star formation locally in the galactic core region (Section \ref{results-sf}).
Newly included radiation feedback from SFMCs, combined with feedback from the MBH, prevents runaway star formation from occurring and helps to retain gas that eventually accretes on to the MBH.
This results in an increased growth rate of the MBH when compared with the run without stellar radiation feedback (Section \ref{results-mbh}).  
As this was the first time that the radiative feedback channels of stars and MBHs are used together in a galaxy-scale simulation,  we are able to demonstrate that previously undiscussed types of interplay between gas, SFMCs, and a MBH may hold important clues about the growth and feedback of quasars and their host galaxies at high $z$. 
We argue that this finding is broadly in line with the previous studies which attempts to find a route to avoid fragmentation of interstellar gas, but to contribute more directly to rapid MBH growth (Section \ref{stability}).  

The comprehensive numerical framework developed for the present study is to overcome the ``mismatch'' between the cutting-edge resolution in the field and the physics models used in simulations.  
Our tests verify that the simulations adopting the proposed numerical framework are computationally feasible on commercially available cluster architectures with sufficiently large memories. 
This opens up numerous possibilities that our numerical framework can be applied to --- including, but not simply limited to, a deeper look into the high-$z$ quasar-host galaxies with even higher resolution.  
Future directions of this line of work are discussed in the next subsection.

\subsection{Future Work}\label{future}

Because we strive to calculate the impact of stellar and MBH feedback energy in the ISM from first principles --- rather than tuning it to match observed galaxies --- and explore previously undiscussed interplay between galactic ingredients, our simulation technique has a potential to offer a unique perspective for the growth of high-$z$ MBHs and their hosts.
Obviously, the present study suffers from various limitations which we will list below. 
We are actively testing and running next generation simulations to expand the scope of our research.

\begin{itemize}

\item Currently, the simulation barely resolves the scale of star-forming gas clumps, and the gas flows at the Bondi radius near a MBH  (see Section \ref{method-mbh}; $\Delta \,x_{14} \, = 4.79 \,\,{\rm pc} \,\,\,{\rm at}\,\,\, z=7.5$).
This was done by employing a strategy to re-simulate an interesting time interval with increased resolution (Section \ref{method-IC}).
A similar technique could be used recursively to further increase resolution between multi-scale re-runs.
Our simulation will be able to follow the actual gas inflows from galaxy- to sub-pc scale in a violent merging event  for which the gas distribution can rarely be idealized as an exponential form --- as sometimes assumed in previous studies such as \cite{2010MNRAS.407.1529H}.\footnote{Note that our re-simulation technique improves that of \cite{2010MNRAS.407.1529H}, one of the most successful attempts to render gas inflows from kpc to sub-pc scale.  To set up a gas distribution for a high-resolution re-simulation, however, they used an idealized, exponential gas disk motivated by the lower-resolution run. In contrast, we plan to adopt the exact gas distribution from the lower-resolution run but with increased resolution.} 
This type of simulations could help us evaluate how galaxy mergers and the triggered MBH activities affect the build-up of high-$z$ quasars, and/or initiate a quasar-regulated star formation phase. 

\item Due to computational limitations, the reported run time is $\sim 25$ Myrs at around $z \sim 7.5$ (for {\it ``Sim-SRTF+BH''}).  
As we advance our simulation for a longer evolution time with better optimizations, we expect to acquire more complete understandings of nonlinear interactions between galactic ingredients occurring at widely different scales simultaneously --- from sub-pc scale to $\gtrsim R_{\rm vir}$ scale.  
For example, we could be poised to study the effect of MBH-driven outflows on gas cooling in the disk and gaseous halo, and if this could alter the star formation history by suppressing cooling and/or gas inflows for a sufficiently long time. 

\item By building a suite of cosmological ``zoom-in'' simulations, we aim to validate the proposed pathways to extremely massive quasars at $z > 7$.
This type of experiments will significantly advance our understandings about the neighborhood of MBHs (such as accretion disk and nuclear disk),  and predict their characteristics. 
While challenging, this is a well-timed study as observations provide excellent constraints on the growths of high-$z$ galaxies and MBHs (Section \ref{intro}). 
More sophisticated physics for MBHs --- absent in the present runs but appropriate at the adopted resolution scale --- should be taken into consideration to refine our estimates:  e.g., MBH accretion model considering angular momentum (not the plain Bondi estimate, Eq.(\ref{eq:Mdot})), panchromatic radiation from a MBH (as opposed to monochromatic one), and outflows from a MBH interacting with magnetic fields \citep{2017ApJ...843..113B}. 
\end{itemize}

\vspace{1mm}

\section*{Acknowledgments}

The authors thank Heon-Young Chang, Mark Krumholz, Myeong-Gu Park, and Hao-Yi Wu for insightful discussions during the progress of this study.
We also thank the anonymous referee who helped significantly improve the draft of this paper.
Ji-hoon Kim acknowledges support by Samsung Science and Technology Foundation under Project Number SSTF-BA1802-04,
and by Research Start-up Fund for the new faculty of Seoul National University. 
John Wise is supported by National Science Foundation grants AST-1614333 and OAC-1835213 and NASA grant NNX17AG23G.
The computing time used for the presented simulation was in part provided by Extreme Science and Engineering Discovery Environment (XSEDE) allocations TG-AST140023 and TG-AST140064.  
XSEDE is supported by the National Science Foundation (NSF) grant ACI-1053575.
Resources supporting this work were also provided by the NASA High-End Computing (HEC) Program through the NASA Advanced Supercomputing (NAS) Division at Ames Research Center.
This work was also supported by the National Institute of Supercomputing and Network/Korea Institute of Science and Technology Information with supercomputing resources including technical support, grants  KSC-2018-S1-0016 and KSC-2018-CRE-0052. 
The publicly available {\sc Enzo} and {\tt yt} codes used in this work are the products of collaborative efforts by many independent scientists from numerous institutions around the world.  
Their commitment to open science has helped make this work possible.

\label{lastpage}

\end{document}